\documentclass[prd,twocolumn,floatfix,preprintnumbers,nofootinbib]{revtex4}
\usepackage[utf8x]{inputenc}
\usepackage{mathrsfs}
\usepackage{amssymb}
\usepackage{amsmath}
\usepackage{mathtools}
\usepackage{graphicx}

\usepackage[caption = false]{subfig}
\usepackage[normalem]{ulem}
\usepackage{xcolor}
\definecolor{lcolor}{rgb}{0.5,0,0}
\definecolor{citcolor}{rgb}{0,0.3,0.0}

\usepackage{array}
\newcommand{\PreserveBackslash}[1]{\let\temp=\\#1\let\\=\temp}
\newcolumntype{C}[1]{>{\PreserveBackslash\centering}p{#1}}
\newcolumntype{R}[1]{>{\PreserveBackslash\raggedleft}p{#1}}
\newcolumntype{L}[1]{>{\PreserveBackslash\raggedright}p{#1}}

\newlength{\mycol}
\usepackage[breaklinks,colorlinks,urlcolor=blue,citecolor=citcolor,linkcolor=lcolor]{hyperref}
\usepackage{mciteplus}

\newcommand{\rt}{{\mathbf{r}}}
\newcommand{\xt}{{\mathbf{x}}}
\newcommand{\xij}[1]{\mathbf{x}_{#1}}
\newcommand{\bt}{{\mathbf{b}}}

\newcommand{\zt}{{\mathbf{z}}}

\newcommand{\ud}{\, \mathrm{d}}

\newcommand{\tr}{\, \mathrm{Tr} \, }

\newcommand{\nc}{{N_\mathrm{c}}}
\newcommand{\nf}{{N_\mathrm{F}}}

\newcommand{\cf}{C_\mathrm{F}}

\newcommand{\nr}[1]{(\ref{#1})}

\newcommand{\gev}{\ \textrm{GeV}}

\newcommand{\lqcd}{\Lambda_{\mathrm{QCD}}}
\newcommand{\as}{\alpha_{\mathrm{s}}}
\newcommand{\asb}{\bar{\alpha}_{\mathrm{s}}}

\newcommand{\fig}{Fig.~}

\newcommand{\eq}{Eq.~}

\newcommand{\xbj}{{x_\text{Bj}}}
\newcommand{\Yobk}{{Y_{0, \text{BK}}}}
\newcommand{\Yoif}{{Y_{0, \text{if}}}}
\newcommand{\etaobk}{{\eta_{0, \text{BK}}}}

\newcommand{\sigmaltnlo}{\sigma_{L,T}^{\textrm{NLO}}}
\newcommand{\sigmaltdip}{\sigma_{L,T}^{\textrm{dip}}}

\newcommand{\sigmaltqgu}{\sigma_{L,T}^{qg, \textrm{unsub.}}}
\newcommand{\sigmalt}[1]{\sigma_{L,T}^{#1}}
\newcommand{\qqbarg}{q \bar{q} g}
\newcommand{\zmin}{z_{2,\textrm{min}}}

\newcommand{\kcal}{\mathcal{K}}
\newcommand{\ocal}{\mathcal{O}}

\newcommand{\der}{\mathrm{d}}

\newcommand{\qso}{Q_\mathrm{s0}}

\begin{document}
\author{G. Beuf}
\affiliation{
Department of Physics, University of Jyv\"askyl\"a %
 P.O. Box 35, 40014 University of Jyv\"askyl\"a, Finland
}
\affiliation{
Helsinki Institute of Physics, P.O. Box 64, 00014 University of Helsinki, Finland
}
\affiliation{
National Centre for Nuclear Research, 02-093, Warsaw, Poland
}
\author{H. Hänninen}
\author{T. Lappi}
\author{H. Mäntysaari}
\affiliation{
Department of Physics, University of Jyv\"askyl\"a %
 P.O. Box 35, 40014 University of Jyv\"askyl\"a, Finland
}
\affiliation{
Helsinki Institute of Physics, P.O. Box 64, 00014 University of Helsinki, Finland
}

\title{
Color Glass Condensate at next-to-leading order meets HERA data
}

%\pacs{}

\preprint{}

\begin{abstract}
We perform the first dipole picture fit to HERA inclusive cross section data using the full next-to-leading order (NLO)  impact factor combined with
an improved Balitsky-Kovchegov evolution including the dominant effects beyond leading logarithmic accuracy at low $x$.
We  find that three different formulations of the evolution equation that have been proposed in the recent literature result in a very similar description of HERA data, and robust predictions for future deep inelastic scattering experiments.
We find evidence pointing towards a significant nonperturbative contribution to the structure function for light quarks, which stresses the need to extend the NLO impact factor calculation to massive quarks.
\end{abstract}

\maketitle

%%%%%%%%%%%%%%%%%%%%%%%%%
\section{Introduction}
%%%%%%%%%%%%%%%%%%%%%%%%%%
The inner structure of protons and nuclei can be accurately determined in deep inelastic scattering (DIS) experiments, where the target structure is probed by a simple pointlike electron via the exchange of a virtual photon. For proton targets, the combined structure function data from the H1 and ZEUS experiments at HERA~\cite{Aaron:2009aa,Abramowicz:2015mha,H1:2018flt,Abramowicz:1900rp} have made it possible to extract the parton densities with an excellent precision. 

At small momentum fraction $x$ the gluon densities rise rapidly, and one eventually expects 
non-linear high-occupancy effects 
to be important and become visible in the weak coupling regime.
At high gluon densities, these non-linear effects tame the growth of the gluon density, and a dynamical scale known as the saturation scale $Q_s^2$ is generated. This scale characterizes the region of phase space where the non-linear saturation effects dominate. To describe QCD in this high energy regime an effective theory known as the Color Glass Condensate has been developed, see Refs.~\cite{Gelis:2010nm,Albacete:2014fwa} for a review. 

The precise DIS data can provide a crucial test for the saturation picture. Theoretically the inclusive DIS cross section is a relatively simple observable, as the probe has no internal structure and one does not need to consider e.g. fragmentation effects. As the proton structure is not perturbatively calculable, some input from experimental data is needed. In the CGC framework, one can calculate the energy 
dependence of various observables, e.g. the total photon-proton cross section, perturbatively by resumming contributions enhanced by a large logarithms of energy or $\ln 1/x$. The non-perturbative input in this case is the proton structure at an initial (and smallish) Bjorken-$x$, which is a parametrized input fitted to the data. The leading order CGC calculations have been able to obtain a good description of the precise HERA data by fitting the initial condition with only a few free parameters~\cite{Albacete:2009fh,Albacete:2010sy,Lappi:2013zma}. However, in all these fits one needs to introduce an additional fit parameter to slow down the $x$-evolution to be compatible with the HERA measurements.

To precisely test the saturation picture of CGC, it is crucial to move beyond leading order accuracy. In recent years the theory has been rapidly developing towards full NLO accuracy. The impact factors, describing the photon-proton interaction, have been calculated at this order in case of massless quarks~\cite{Balitsky:2010ze,Balitsky:2012bs,Beuf:2011xd,Beuf:2016wdz,Beuf:2017bpd, Hanninen:2017ddy}, and  the first numerical results were reported in Ref.~\cite{Ducloue:2017ftk}. The impact factors need to be combined with evolution equations that describe the Bjorken-$x$ dependence and resum contributions enhanced by large logarithms of energy, $(\as \ln 1/x)^n$ at leading order and $\alpha_s(\alpha_s \ln 1/x)^n$ at next-to-leading order. The Balitsky-Kovchegov (BK) equation describing the evolution of the dipole-target interaction~\cite{Balitsky:1995ub,Kovchegov:1999yj} is available at NLO accuracy~\cite{Balitsky:2008zza} with the higher order contributions enhanced by large transverse logarithms resummed in Refs.~\cite{Beuf:2014uia,Iancu:2015vea,Iancu:2015joa,Ducloue:2019ezk} and  numerical solutions reported in Refs.~\cite{Lappi:2015fma,Lappi:2016fmu}.

An additional complication in the small-$x$ evolution is that the Coulomb tails obtained from a perturbative calculation result in the proton size growing much faster than seen in the data, and faster than suggested by the Froissart bound~\cite{Froissart:1961ux} for hadronic collisions. It has been argued~\cite{Kovner:2001bh,Kovner:2002yt} on the theoretical level that including some non-perturbative damping of the gluon emission at large transverse distance is necessary and sufficient to recover a Froissart behavior for the virtual photon-proton cross section. This idea has been studied~\cite{Berger:2011ew,Bendova:2019psy} in fits to the HERA data using the impact parameter dependent BK equation supplemented by either a non-perturbative cut-off or collinear resummations.
In addition to the BK equation, one can solve the more general JIMWLK evolution equation~ \cite{JalilianMarian:1996xn,JalilianMarian:1997jx, JalilianMarian:1997gr,Iancu:2001md, Ferreiro:2001qy, Iancu:2001ad, Iancu:2000hn} (available at NLO accuracy~\cite{Balitsky:2013fea,Kovner:2013ona}, but no numerical solution exists for the NLO equation). The JIMWLK evolved proton structure was compared with the HERA data in Ref.~\cite{Mantysaari:2018zdd} (see also Refs.~\cite{Schlichting:2014ipa,Mantysaari:2019jhh}), where again large non-perturbative contributions were needed to describe the system with a finite proton geometry. Due to these additional complications, we only study an impact parameter independent evolution here, and assume that the transverse area of the proton can be factorized in the cross section calculations.

In addition to testing the saturation conjecture, an accurate description of the DIS data is important for other phenomenological applications. As we will discuss later, the DIS cross section is written in terms of the quark dipole-target scattering amplitude. The exactly same degrees of freedom are needed to describe other scattering processes, such as particle production in proton-nucleus collisions (see e.g.~\cite{Albacete:2010bs,Lappi:2013zma,Fujii:2013gxa,Ducloue:2015gfa,Ducloue:2016pqr,Ducloue:2016ywt,Ma:2014mri,Ma:2017rsu,Mantysaari:2019nnt,Tribedy:2010ab,Lappi:2012nh,Ducloue:2017kkq}) or diffractive DIS (e.g.~\cite{Marquet:2007nf,Lappi:2010dd,Lappi:2013am,Mantysaari:2016ykx,Mantysaari:2016jaz,Mantysaari:2017slo,Salazar:2019ncp,Mantysaari:2019csc,Mantysaari:2019hkq}). Although most of the current phenomenological applications are performed at leading order accuracy, the NLO calculations are developing rapidly~\cite{Chirilli:2012jd,Stasto:2013cha,Altinoluk:2014eka,Altinoluk:2015vax,Ducloue:2016shw,Boussarie:2016ogo,Boussarie:2016bkq,Ducloue:2017dit,Liu:2019iml,Escobedo:2019bxn,Roy:2019hwr,Liu:2020mpy}. A necessary input for the phenomenological applications at NLO accuracy is the initial condition for the NLO evolution, which can be obtained by fitting the DIS data as presented in this paper.

This paper is structured as follows. First, in Sec.~\ref{sec:dis}, we will briefly introduce the dipole picture of DIS at leading and  next-to-leading order. Then, in Sec.~\ref{sec:evolution}, we will review the necessary details of the different variants of the BK equation used in this work. Section~\ref{sec:data} reviews the data sets used in the fits, and Sec.~\ref{sec:fitresults} discusses the results of the fits.

\section{Deep inelastic scattering in the dipole picture at NLO}
\label{sec:dis}
The  photon-proton cross section is parametrized in terms of the structure functions $F_2$ and $F_L$, that are related to the virtual photon-proton cross sections $\sigma^{\gamma^*p}$ as
\begin{equation}
    F_2(\xbj,Q^2) = \frac{Q^2}{4\pi^2 \alpha_\text{em}} \left(\sigma_L^{\gamma^* p} + \sigma_T^{\gamma^* p} \right)
\end{equation}
and
\begin{equation}
    F_L(\xbj,Q^2) = \frac{Q^2}{4\pi^2 \alpha_\text{em}} \sigma_L^{\gamma^* p} .
\end{equation}
Here the subscripts $T$ and $L$ refer to the transverse and longitudinal polarizations of the virtual photon. The experimental data is often reported as a reduced cross section 
\begin{multline}
    \sigma_r(\xbj,y,Q^2) = F_2(\xbj,Q^2)
    \\
    - \frac{y^2}{1 + (1-y)^2} F_L(\xbj,Q^2).
\end{multline}
Here $-Q^2$ is the photon virtuality, $\xbj$ is the Bjorken variable and $y$ is the inelasticity.  

The focus in this paper is on the next-to-leading order corrections to the total DIS cross section in the dipole picture. As an introduction, let us first briefly describe the process in the  leading order dipole picture.

At leading order, the virtual photon-proton scattering in the dipole picture is understood in the following way (see e.g.~\cite{Kovchegov:2012mbw}). First, the incoming photon fluctuates into a quark-antiquark pair. This splitting is described by the photon light cone wave function $\psi^{\gamma^*\to q\bar q}$. Subsequently, the produced dipole interacts with the target. At high energy, the quark-target interaction is eikonal, and the transverse position of the quark does not change during the scattering. Instead, the quark goes through a color rotation in the target color field and picks up a Wilson line $V(\xij{0})$ in the fundamental representation, where $\xij{0}$ is the transverse coordinate of the quark. Similarly, the antiquark at point $\xij{1}$ picks up a conjugate Wilson line $V^\dagger(\xij{1})$. 

To calculate the total cross section, one applies the optical theorem and calculates the imaginary part of the forward elastic scattering amplitude for the process $\gamma^* p \to \gamma^* p$. The resulting cross section reads
\begin{multline}
\label{eq:lo-cross-section}
    \sigma^{\gamma^* p}_{T,L}=2 \int \der^2 \bt \der^2 \rt \der z |\psi^{\gamma^* \to q\bar q}(\rt,Q^2,z)|^2
    \\
    \times
    \left(1-S(\rt, \bt, x)\right).
\end{multline}
Here, $z$ is the light-cone 
momentum fraction of the photon carried by the quark. 
The dipole size is $\rt=\xij{0}-\xij{1}$ and its
impact parameter is $\bt=(\xij{0}+\xij{1})/2$.
In the following, $S(\rt, \bt, x)$ is assumed to depend only slowly on $\bt$. Thus we will drop this dependence on $\bt$ and replace the integration over $\bt$ by a constant $\int \der^2 \bt \to \sigma_0/2$. 
The dipole scattering matrix $S$ is defined as a two point function of the Wilson lines that the quarks pick up in the scattering process:
\begin{equation}
\label{eq:dipole}
    S(\rt, \bt, x) \equiv
        \left\langle 
            \frac{1}{N_c}
            \tr V(\xij{0})V^{\dagger}(\xij{1})
        \right\rangle_{x}.
\end{equation}
The brackets $\langle \rangle$ refer to the average over the target color charge configurations. Here the momentum fraction $x$ in the subscript stands for the fact that the  Wilson lines are evaluated  at some energy or rapidity scale corresponding to the kinematics of the process. This dependence is given by the Balitsky-Kovchegov equation which, at leading order, is usually used to evolve the Wilson lines up to an evolution rapidity $Y = \log 1/\xbj$. At NLO the question of the evolution rapidity becomes more complicated, as discussed in more detail in Sec.~\ref{sec:evolution}. 

At Next-to-Leading Order (NLO) the virtual photon-proton scattering involves Fock states of the photon that contain a gluon in addition to the quark and antiquark, which all scatter off the target. There are also other NLO contributions with only a quark-antiquark Fock state scattering off the target, which include a gluon loop correction to the photon splitting.
 These NLO $q\bar{q}g$ and $q\bar{q}$ contributions have been calculated independently using the conventional dimensional regularization~\cite{Beuf:2016wdz, Beuf:2017bpd} and four dimensional helicity schemes~\cite{Hanninen:2017ddy}. The individual diagrams contain UV divergences that cancel each other in the sum.
On top of these, there remains a divergence related to low $x$ gluons, 
which must be resummed 
into the evolution of the target. Subtraction schemes for this low $x$ gluon divergence in DIS were devised and tested in Refs.~\cite{Beuf:2017bpd,Ducloue:2017ftk}, and in our present paper we continue to refine the `unsub' scheme to enable a comparison between the theory and experimental data.

In Refs.~\cite{Beuf:2017bpd,Ducloue:2017ftk} the low $x$ gluon divergence factorization from the NLO DIS cross sections (for a more detailed discussion in the context of single inclusive particle production see Refs.~\cite{Ducloue:2016shw,Iancu:2016vyg,Ducloue:2017mpb}) were written in two distinct but equivalent forms: a form where the factorization is implicit, and another where it was made explicit, named `unsubtracted' and `subtracted' schemes respectively. In this work we use the unsubtracted form for the cross sections, which can be expressed as
\begin{equation}
    \sigmaltnlo = \sigmalt{\textrm{IC}} + \sigmaltdip + \sigmaltqgu .
\end{equation}
Here the first term is the leading order cross section \eqref{eq:lo-cross-section} where the dipole scattering amplitude is evaluated at the chosen fixed initial rapidity scale of the target, corresponding to the initial condition of BK evolution. The other terms can be interpreted as arising from the NLO $q\bar{q}$ diagrams ($\sigmaltdip$)
 and from the NLO $q\bar{q}g$ diagrams ($\sigmaltqgu$), up to subtraction terms used to make the cancellation of UV divergences between these diagrams explicit.
In our scheme, the 'unsubtracted' $qg$ term is:
\begin{align}
    \sigmaltqgu & = 
        \notag
        8 \nc \alpha_\text{em} \frac{\alpha_s \cf}{\pi} \sum_f e_f^2
        \int_0^1 \der z_1 \int_{\zmin}^{1-z_1} \frac{\der z_2}{z_2}
        \\ \label{eq:NLO_qg_unsub}
        &
        \times
        \mkern-18mu % quad ~ mkern-18mu
        \int\displaylimits_{\xt_0, \xt_1, \xt_2}
        \mkern-18mu % quad ~ mkern-18mu
        \mathcal{K}_{L,T}^{\textrm{NLO}}(z_1, z_2, \xt_0, \xt_1, \xt_2),
\end{align}
and the 'dipole' term is:
\begin{align}
     \sigmaltdip
        &= 4 \nc \alpha_{em} \frac{\alpha_s \cf}{\pi} \sum_f e_f^2 \int_0^1 \ud z_1
        \nonumber
        \\ \label{eq:NLO_dip}
        &
        % \!\!\!
        \times
        \mkern-12mu
        \int\displaylimits_{\xt_0, \xt_1} 
        \mkern-9mu % quad ~ mkern-18mu
        \kcal_{L,T}^\text{LO}
        \!
        \left(z_1,\xt_0,\xt_1\right) 
        \!\!
        \left[\frac{1}{2}\ln^2\!\left(\!\frac{z_1}{1\!-\!z_1}\!\right)\!-\!\frac{\pi^2}{6}\!+\!\frac{5}{2}\right],
\end{align}
with the shorthand 
% \begin{equation}
$\int_{\xt_i} \coloneqq \int \frac{\der^2 \xt_i}{2 \pi}$.
% \end{equation}
The integrand kernels and dipole operators for the leading order and 'dipole' terms are
\begin{eqnarray}
    \kcal_{L}^\text{LO}(z_1,\xt_0,\xt_1) 
        & =& 4 Q^2 z_1^2 (1-z_1)^2
        \nonumber \\
        & & \quad 
        \times 
        K_0^2(Q X_2) \left(1-S(\xij{01})\right),
    \\[2ex]
    \kcal_{T}^\text{LO}(z_1,\xt_0,\xt_1)
        & =& Q^2 z_1 (1-z_1) \left(z_1^2+(1-z_1)^2\right)
        \nonumber \\
        & & \quad
        \times
        K_1^2(Q X_2) \left(1-S(\xij{01})\right),
\end{eqnarray}
where $X_2^2 \equiv z_1 ( 1 - z_1 ) \xt_{01}^2$, $\xt_{ij} \equiv \xt_i - \xt_j$ and $S(\xt_{ij}) \equiv S(\xt_{ij},\bt)$.
Here, the rapidity scale which the dipole operator \eqref{eq:dipole} is evaluated at is left implicit. It will be discussed together with the associated small-$x$ evolution in Sec.~\ref{sec:evolution}. However, we note already now that in the $qg$-term this rapidity scale must be taken to depend on the gluon momentum fraction $z_2$, not just the external kinematical scales $\xbj$ and $Q^2$. This is essential for the stability of the factorization scheme, as discussed in great detail e.g. in  Refs.~\cite{Beuf:2014uia,Ducloue:2016shw,Iancu:2016vyg,Ducloue:2017mpb,Ducloue:2017ftk}.  

For the $qg$-terms that only appear at NLO, the kernels and Wilson line operators are:
\begin{widetext}
    \begin{align}
        &\kcal_L^\text{NLO}(z_1,z_2,\xt_0,\xt_1,\xt_2)= 4 Q^2 z_1^2 (1-z_1)^2 \bigg\{ \! P \! \left(\frac{z_2}{1-z_1}\right) \! \frac{\xt_{20}}{\xt_{20}^2} \!\cdot\! \left(\frac{\xt_{20}}{\xt_{20}^2}-\frac{\xt_{21}}{\xt_{21}^2}\right) \! \left[ K_0^2(Q X_3) \left(1-S_{012}\right)-(\xt_2 \to \xt_0) \right] \nonumber \\
        & \hspace{7cm}+ \left(\frac{z_2}{1-z_1}\right)^2 \frac{\xt_{20} \cdot \xt_{21}}{\xt_{20}^2 \xt_{21}^2} K_0^2(Q X_3) \left(1-S_{012}\right) \bigg\} ,
        \\
        &\kcal_T^\text{NLO}(z_1,z_2,\xt_0,\xt_1,\xt_2)= Q^2 z_1(1-z_1) \nonumber \\
        &\hspace{1cm}\times \bigg\{ 
        P\left(\frac{z_2}{1-z_1}\right)\left(z_1^2+(1-z_1)^2\right) \frac{\xt_{20}}{\xt_{20}^2} \cdot \left(\frac{\xt_{20}}{\xt_{20}^2}-\frac{\xt_{21}}{\xt_{21}^2}\right) \Big[K_1^2(Q X_3) \left(1-S_{012}\right)-(\xt_2 \to \xt_0) \Big] \nonumber \\
        &\hspace{1.6cm} + \left(\frac{z_2}{1-z_1}\right)^2 \left[ \left(z_1^2+(1-z_1)^2\right) \frac{\xt_{20} \cdot \xt_{21}}{\xt_{20}^2 \xt_{21}^2} + 2 z_0 z_1 \frac{\xt_{20} \cdot \xt_{21}}{\xt_{20}^2 X_3^2} - \frac{z_0(z_1+z_2)}{X_3^2} \right] K_1^2(Q X_3) \left(1-S_{012}\right) \bigg\} .
    \end{align}
\end{widetext}
Here $z_0, z_1, z_2$ are the longitudinal momentum fractions of the quark, antiquark and gluon, which satisfy ${\sum_i z_i = 1}$. The parton configuration factor $QX_3$ is interpreted as the ratio of the $\qqbarg$ state formation time to the $\gamma^{*}$ lifetime~\cite{Beuf:2011xd}. It is defined as
${ X_3^2 \coloneqq z_0 z_1 \xt_{01}^2 + z_0 z_2 \xt_{02}^2 + z_2 z_1 \xt_{21}^2 }$. % braces make this a 'math atom' that doesn't line break.
We have also defined a shorthand  $ P(z) \coloneqq 1 + (1-z)^2$. The $\qqbarg$ state-target scattering Wilson line operator is 
\begin{equation}
    S_{012} \equiv \frac{\nc}{2\cf}
        \left(
            S(\xij{02})S(\xij{21}) - \frac{1}{\nc^2}S(\xij{01})
        \right).
\end{equation}
In \eq \eqref{eq:NLO_qg_unsub} the lower limit $z_{2,\text{min}}$ in the gluon longitudinal momentum fraction integral is yet undefined, and its proper value will be discussed in the next section.

%%%%%%%%%%%%%%%%%%%%%%%%%%%%%%%%%%%%%%%%%%%%%%%%
\section{High energy evolution }
\label{sec:evolution}
%%%%%%%%%%%%%%%%%%%%%%%%%%%%%%

\subsection{Balitsky-Kovchegov equation}
\label{sec:bkgeneral}

In the calculation of the photon-proton cross section at NLO, as discussed above, the dipole-target scattering amplitude depends on the energy, or equivalently, on Bjorken-$x$. In the large $\nc$ limit, the evolution is given by the Balitsky-Kovchegov (BK) equation~\cite{Balitsky:1995ub,Kovchegov:1999yj}. At leading order, the BK equation reads
\begin{multline}
    \label{eq:bk-evolution}
    \frac{\partial S(\xij{01})}{\partial Y} =  \int \der^2 \xij{2} K_\text{BK}
    (\xij{0}, \xij{1}, \xij{2})
    \\
    \times
    [S(\xij{02}) S(\xij{21}) - S(\xij{01})].
\end{multline}
The kernel $K_\text{BK}$ is proportional to the probability density to emit a gluon with transverse coordinate $\xij{2}$ from the dipole of size $\xij{01}=\xij{0}-\xij{1}$. The evolution rapidity $Y$ is discussed in detail later. When  running coupling corrections following the Balitsky prescription~\cite{Balitsky:2006wa} are included, it reads
\begin{multline}
\label{eq:bk-rc-balitsky}
  K_\text{BK}(\xij{0},\xij{1}, \xij{2}) = \frac{\nc \as(\xij{01}^2)}{2\pi^2} \left[
        \frac{\xij{01}^2}{\xij{12}^2 \xij{02}^2} \right. \\
        + \frac{1}{\xij{02}^2} \left( \frac{\as(\xij{02}^2)}{\as(\xij{12}^2)} -1 \right) 
        + \left. \frac{1}{\xij{12}^2} \left( \frac{\as(\xij{12}^2)}{\as(\xij{02}^2)} -1 \right)
  \right].
\end{multline}

In principle we should use the next-to-leading order BK equation when using the impact factors calculated to the order $\as$. 
The required numerical solution of the NLO BK equation exists~\cite{Lappi:2015fma,Lappi:2016fmu,Lappi:2020srm}. However, the equation is numerically burdensome due to the high dimensional transverse integration (in the NLO BK equation one integrates over the transverse coordinates of the two emitted gluons, instead of just one gluon in the leading order equation). Instead of the full equation, in this work we use prescriptions of BK evolution that capture an important subset of beyond leading order effects. The difference between the studied evolutions reflects some of  the uncertainty due to the missing full NLO evolution.

In practice we have chosen three related formulations of the BK equation that resum some or all of the large transverse momentum logarithms in the NLO equation. Firstly we consider the nonlocal evolution equation in terms of the projectile momentum fraction introduced in Ref.~\cite{Beuf:2014uia}, where collinear double logarithms are resummed via the inclusion of  a Kinematical Constraint: we denote this the KCBK equation. Secondly, we consider the local equation in the projectile momentum fraction of Ref.~\cite{Iancu:2015vea}, where the same double logarithms, together with DGLAP-like single logarithms, are explicitly resummed into a kernel that is a nontrivial function of $\as$: we call this the ResumBK equation. Thirdly we study a  nonlocal equation in the target momentum fraction, recently formulated in Ref.~ \cite{Ducloue:2019ezk} and denoted here as the TBK equation.
The first two are formulated in terms of the  projectile momentum fraction, so that the rapidity-like evolution variable in the BK equation is defined by the momentum fraction of the probe, or equivalently by the plus component of the 4-momentum\footnote{We work in a frame where the target has a large minus momentum $P^-$, and the incoming photon has a large plus momentum $q^+$.}.
Since the projectile momentum fraction is the variable appearing explicitly in   the NLO DIS impact factors,  using these evolution equations is fairly straightforward. However the fact that the TBK equation  is written in terms of the target momentum fraction, i.e. the minus component of 4-momentum,  means that the  evolution and the perturbative impact factors can only be matched approximatively, and the procedure requires more care. 

As we will discuss in detail in the following, we solve the BK equation  by taking a parametrized dipole amplitude as an initial condition at an initial rapidity scale. Starting from  this initial condition the evolution predicts the behavior of the dipole at higher rapidities, i.e. energies, or correspondingly at smaller Bjorken-$x$.
The phase space available for the emission of the gluon grows with energy and determines the amount of BK evolution. Thus
the evolution range is controlled by the lower limit of the gluon momentum  fraction  $\zmin$ in Eq.~\eqref{eq:NLO_qg_unsub}, with a smaller lower limit corresponding to longer evolution.

\subsection{Evolution in projectile momentum fraction}
\label{sec:y}

Let us first consider the evolution written in projectile momentum fraction, which is the case for the KCBK and ResumBK equations. In this case the high energy evolution is parametrized by the  rapidity variable $Y$, which is defined using the plus components of the gluon momentum $k^+$ and a plus momentum scale $P^+$  associated with the target as 
\begin{equation}
Y \equiv \ln\left(\frac{k^+}{P^+}\right).
\end{equation}
Since the incoming photon energy $q^+$ (which is the maximal $k^+$) in the target rest frame is proportional to the photon-target c.m.s. energy $W^2$, one should think of evolution in the rapidity variable $Y$ as evolution in $\ln W^2$, as we will see more explicitly below.

In the impact factor, Eq.~\eqref{eq:NLO_qg_unsub}, the  gluon momentum is parametrized by the momentum fraction $z_2$ as $k^+ = z_2 q^+$. Both the probe momentum fraction evolution equations and the  NLO impact factor are derived in terms of the same $z_2$. Thus it is straighforward to see that the dipole operators in the evaluation of the $qg$-term in the cross section~\eqref{eq:NLO_qg_unsub} are always evaluated at the projectile rapidity 
\begin{equation} 
Y=\ln z_2 + \ln\left(\frac{q^+}{P^+}\right),
\end{equation}
depending on the integration variable $z_2$. We will specify the value of $P^+$ below.

First, we have to determine the lower limit $z_{2,\text{min}}$ for the $z_2$ integral in the NLO impact factor, Eq.~\eqref{eq:NLO_qg_unsub}, which controls the amount of evolution.
This limit is set by the overall kinematics of the process. One way to understand the existence of this limit is to note that in the limit $z_2\to 0$ the invariant mass of the $q\bar{q}g$-system interacting with the target grows as $M^2_{q\bar{q}g} \sim 1/z_2$. The fact that this invariant mass cannot be larger than the c.m.s. collision energy results in a lower limit for kinematically allowed values of $z_2$. Since the validity of the eikonal approximation used to derive the dipole picture cross section requires in principle $M^2_{q\bar{q}g} \ll W^2$, one could require a more strict limit on $z_2$ than resulting from purely kinematics. Thus there is a choice in how close to the kinematical limit one allows the integral to go, which we quantify by the parameter  $e^{\Yoif} \gtrsim 1$. In terms of this parameter we have the limit
\begin{equation}
\label{eq:z2min}
\begin{split}
    z_2 q^+ &> e^{\Yoif} P^+ = e^{\Yoif} \frac{Q_0^2 }{2P^-} = e^{\Yoif} \xbj \frac{Q_0^2}{Q^2} q^+ \\
    z_2 &> e^{\Yoif} \xbj \frac{Q_0^2}{Q^2}\approx e^{\Yoif}  \frac{Q_0^2}{W^2} \equiv \zmin.
    \end{split}
\end{equation}
Here we have introduced a non-perturbative target transverse momentum scale $Q_0^2$, for which in this work we use the value\footnote{Note that the two parameters $\Yoif$ and $Q_0$ only appear in one combination $e^{\Yoif} Q_0^2$ here; thus there is really only one independent parameter characterizing the limit $\zmin$. However, for the discussion that follows it is better to think in terms of a separate nonperturbative transverse momentum scale $Q_0$. } $Q_0^2= 1\gev^2$. This allows us to write $P^+= Q_0^2/(2P^-)$, and we used the fact that $\xbj={Q^2/(2 P\cdot q) = Q^2/(2 P^- q^+)}$. This limit is already derived e.g. in Refs.~\cite{Beuf:2014uia,Beuf:2017bpd,Ducloue:2017ftk}. In Ref.~\cite{Ducloue:2017ftk} the authors for simplicity set $Q_0^2/Q^2=1$ in practical evaluations of the NLO impact factors. 

In principle also the limits $z_1\to 0$, $z_1 \to 1$ in Eqs.~\eqref{eq:NLO_qg_unsub} and~\eqref{eq:NLO_dip} correspond to the invariant mass of the scattering state becoming infinite, similarly to the limit $z_2 \to 0$. Thus, as discussed in Ref.~\cite{Beuf:2017bpd}, one could also take the energy or rapidity scale at which the dipoles are evaluated to depend on the (anti) quark momentum fractions $z_0,z_1$ (see also Ref.~\cite{Bialas:2000xs}). This part of phase space does not, however, generate a   contribution enhanced by a large logarithm of $x$ to the cross section. Instead, this ``aligned jet'' configuration produces a large collinear logarithm which in principle should be included in the DGLAP evolution not included in the dipole picture applied in this work. Properly including this collinear logarithm is an important issue but separate from the factorization to the BK equation, and is left for future work.
%Thus we leave it to further work to explore improved treatments of these ``aligned jet'' contributions.

When the Bjorken-$x$ of the process is such that the smallest momentum fraction $\zmin$ is close to $1$, i.e. when $\xbj \sim e^{-\Yoif}$ (with $Q^2 \sim Q_0^2$), the possible phase space for real gluon emission allowed in the expression for the cross section vanishes. Thus the $qg$ contribution to the NLO cross section goes to zero at $\xbj \sim e^{-\Yoif}$ by construction. 
The NLO calculation does not fix any exact value for $\Yoif$. A possible choice to consider for $\Yoif$
would be  to take $\Yoif \approx \ln 1/0.01 $, corresponding to the limit were the dipole picture is usually considered applicable.  This was the choice used in  Ref.~\cite{Ducloue:2017ftk}. However, this choice leads to a transient effect in the NLO cross sections at the upper end of the $\xbj$ range $\xbj \sim e^{-\Yoif}$ 
since the positive virtual correction remains large, while the  the negative $qg$ contribution vanishes, as demonstrated in Ref.~\cite{Ducloue:2017ftk}. 

To avoid this unphysical transient effect, we adopt here instead the maximal (or minimal depending on the point of view) choice  $\Yoif=0$. This means that the integral over $z_2$ in the cross section extends all the way to the kinematical limit, outside of the validity of the eikonal approximation. The contribution from this region is, however, only a parametrically small part of the cross section for small $\xbj$, which is where we are comparing the cross section to experimental data. Also, since there is a cancellation between the real and virtual contributions to the cross section, and the latter includes a $z_2$ integral over the full range $0<z_2<1$, one could in fact argue that this choice minimizes the net effect of very large invariant mass states in the photon on the cross section. 

The above discussion only applies to the $qg$ term~\eqref{eq:NLO_qg_unsub} in the cross section.  The virtual correction in Eq.~\eqref{eq:NLO_dip} is already integrated over $z_2$ and cannot be evaluated at a $z_2$-dependent rapidity. Thus for this term dipole operators are taken to be independent of $z_2$ and evaluated at rapidity $Y=\ln 1/\xbj$. Using a $z_2$ independent dipole is justified, as the region $z_2 \ll 1$ gives only a negligible contribution to the virtual correction. Including these formally subleading effects, namely the $z_2$ dependent dipole operator in the virtual term and improving the approximation $Y\approx \ln 1/\xbj$ is left for future work.

The choice $\Yoif=0$ removes the unphysical transient effect, but it forces us to confront another problem that the earlier formulation of Ref.~\cite{Ducloue:2017ftk} wanted to avoid by choosing a larger $\Yoif$. Namely, at the lower end of the $z_2$ range we are forced to evaluate also the (BK-evolved) dipoles at a rapidity scale that is lower (or $\xbj$-scale that is higher) than where the BK-equation is normally used. Now we again have different options regarding the rapidity where we start the  the BK evolution. We parametrize this choice by another constant $\Yobk$, whose value can also be chosen in different ways.

One way is to take $\Yobk = \Yoif=0$, in which case we simply start the BK evolution much earlier (much higher $\xbj$) than where we are actually calculating the cross section. 
Here the contribution of the unphysical small rapidity or large $x$ phase space to the cross section~\nr{eq:NLO_qg_unsub} is suppressed, because  target gets more and more dilute following the evolution backwards to smaller rapidities.
This procedure changes the way the parametrization of the initial condition for the BK-evolution should be interpreted. In this approach, the quantity that can meaningfully be compared to the initial dipole amplitude at $x=x_0\sim 0.01$ in LO fits is not the actual initial condition at $Y=\Yobk=0$, but the result obtained after $Y=\ln 1/0.01$ units of rapidity evolution.

Another option is to take a more typical initial energy scale for the BK equation, which we here take as $\Yobk = \ln 1/0.01 $. In this latter case one has to model the dipole amplitude in the region $\Yoif < Y < \Yobk$.  In this case, we simply assume that the dipole operator is independent of $Y$ in this region, which is ``before the initial condition'' in $Y$. Assuming an energy independent dipole amplitude in this region  is consistent within the accuracy of the framework.

To summarize, we have two parameters that we must choose, $\Yoif$ and $\Yobk$. In this work we always take  $\Yoif=0$ to avoid the large transient effect at $\xbj\sim e^{-\Yoif}$ in the data region. We then  apply two approaches for the parameter  $\Yobk$. The first option is to start the BK evolution at rapidity $\Yobk = \ln 1/0.01$ and freeze the dipole amplitude at $Y < \Yobk$.  The second option is to  also start the BK evolution at rapidity $\Yobk=\Yoif=0$.

We recall that the dipole amplitudes in the cross section  \eq \eqref{eq:NLO_qg_unsub} are evaluated at a rapidity 
\begin{equation}
\label{eq:y_z2}
Y \equiv \ln \frac{k^+}{P^+} = \ln \frac{W^2 z_2}{Q_0^2}.
\end{equation}
Note that the maximum  rapidity $Y_\text{max}$ encountered is obtained at the $z_2\to 1$ limit
\begin{equation}
    \label{eq:y-max}
    Y_\text{max} = \ln \frac{W^2}{Q_0^2} = \ln \frac{1}{\zmin} + \Yoif,
\end{equation}
corresponding to values of $Y$ probed by the  the $z_2$-integral in the cross section ranging from $\Yoif$ to $Y_\text{max}$, i.e. over a rapidity interval $\Delta Y = \ln 1/\zmin $. 
When $\Yobk > \Yoif$, the actual range of BK evolution is smaller by $\Yobk$, and for $\Yoif< Y < \Yobk$ the dipole does not change.
We emphasize that the evolution rapidity depends only on the total center-of-mass energy $W^2$, and not explicitly on $\xbj$ or $Q^2$. This is natural, as the scattering amplitude for a dipole with a fixed transverse size can only be sensitive to the total center-of-mass energy, and strictly speaking the dipole does not exactly know about the photon virtuality or the Bjorken $x$.

\subsection{Kinematically constrained BK}
\label{sec:kcbk}

As discussed previously, in order to keep the computational cost of the fit procedure manageable, we do not use the full NLO BK equation to obtain the rapidity dependence of the dipole scattering matrix $S$. Instead, we use modified versions of the leading order evolution equation that
resums the most important higher order corrections, in particular the collinear double logarithms.

First, we use the KCBK equation~\cite{Beuf:2014uia} that is nonlocal in the projectile momentum fraction (i.e. the evolution variable $Y$):
\begin{align}
\label{eq:kcbk}
    & \partial_Y S(\xij{01},Y) =
    \notag
    \\
    & \, \int \der^2\xij{2}
    K_\text{BK}(\xij{0}, \xij{1}, \xij{2})
        \theta\left(Y - \Delta_{012} - \Yoif \right)
    \notag
    \\
    & \, \, \times \! \left[ S(\xij{02}, Y \! - \! \Delta_{012}) S(\xij{21}, Y \! -\! \Delta_{012})- S(\xij{01}, Y)  \right]
\end{align}
with
\begin{equation}
    \label{eq:kcbk-shift}
    \Delta_{012} = \max \left\lbrace 0, \ln \frac{\min\{ \xt_{02}^2,\xt_{21}^2 \} }{\xt_{01}^2} \right\rbrace.
\end{equation}
This equation explicitly forces time ordering between subsequent gluon emissions.  The theta function ensures that only dipoles in the range $Y>\Yoif$ are included.

\subsection{Rapidity local resummed BK}
\label{sec:resum}

The most important higher order corrections to the BK equation that are enhanced by double large transverse logarithms 
can be resummed alternatively into a kernel that is local in the evolution rapidity $Y$ by a method introduced in Ref.~\cite{Iancu:2015vea}\footnote{The double log resummation was further developed in Ref.~\cite{Ducloue:2019ezk}, in this work we however use the result from Ref.~\cite{Iancu:2015vea} numerically implemented in Ref.~\cite{Lappi:2016fmu}.}. This procedure resums exactly the same contributions that are included in Ref.~\cite{Beuf:2014uia} to derive the kinematically constrained BK equation shown above in Eq.~\eqref{eq:kcbk}. A practical advantage of the approach taken in Ref.~\cite{Iancu:2015vea} is that the resulting equation is local in evolution (projectile) rapidity, and as such numerically easier to solve using standard Runge-Kutta methods. 
In addition to the double transverse logarithms resummation, the contribution of some of the single transverse logarithms present in the NLO BK equation can be included following Ref.~\cite{Iancu:2015joa}, keeping the equation local in rapidity.
In Ref.~\cite{Albacete:2015xza} it was shown that this resummed BK equation is in practice  close to the kinematically constrained BK equation discussed previously. 
As the resulting resummed evolution equation is written in terms of the projectile rapidity $Y$, it can be used with the impact factors exactly as the kinematically constrained BK equation.

The resummed equation is obtained by multiplying the BK kernel \eqref{eq:bk-rc-balitsky} by $K_\text{DLA} K_\text{STL}$, where $K_\text{DLA}$ is a resummation of double and $K_\text{STL}$ single transverse logarithms.  The kernel resumming the double transverse logarithms reads
\begin{equation}
    \label{eq:resum-dla}
    K_\text{DLA} = \frac{J_1(2 \sqrt{\bar \as x^2})}{\sqrt{\bar \as x^2}},
\end{equation}
with $x=\sqrt{ \ln \xij{02}^2/\xij{01}^2 \ln \xij{12}^2/\xij{01}^2}$ and $\bar \as = \as \nc/\pi$. If $ \ln \xij{02}^2/\xij{01}^2 \ln \xij{12}^2/\xij{01}^2< 0$, an absolute value of the argument is used and the Bessel function is changed to $J_1 \to I_1$, see Ref.~\cite{Iancu:2015vea}.
The single transverse logarithms $\sim \as \ln 1/(\xij{ij}^2 Q_s^2)$  are included  multiplying the kernel  multiplied by
\begin{equation}
    \label{eq:resum-stl}
    K_\text{STL} = \exp \left\{ -\frac{\as \nc A_1}{\pi} \left| \ln \frac{C_\text{sub} \xij{01}^2}{\text{min}\{\xij{02}^2, \xij{12}^2\}}  \right| \right\}.
\end{equation}
In Ref.~\cite{Lappi:2016fmu} it was shown that the resummation of single transverse logarithms can be done such that the resummed equation is a good approximation to the full NLO BK evolution by adjusting the constant $C_\text{sub}$ whose numerical value is not fixed by the resummation procedure. This renders the $\mathcal{O}(\as^2)$ contributions in the NLO BK equation that are not enhanced by large (single) transverse logarithms minimal. With this procedure, one obtains a rapidity local projectile momentum fraction resummed BK equation which we use as an approximation to the full NLO BK equation (with a resummation of large transverse logarithmic corrections), with $C_\text{sub}=0.65$ determined in Ref.~\cite{Lappi:2016fmu}. 

The resummation of the single transverse logarithms is completely independent of the resummation of the double transverse logarithms, and thus could be included in the same way also in the other studied evolution equations. In this work, however, we only include this contribution in the ResumBK evolution, as we prefer to work with the established versions of the BK evolution. We will discuss the effect of the single transverse logarithm resummation on our fits in Sec.~\ref{sec:fitresults}.

\subsection{Target momentum fraction evolution}
\label{sec:trbk}

As discussed in detail in Ref.~\cite{Ducloue:2019ezk}, it is possible to formulate the evolution in terms of the target rapidity $\eta$ defined as a logarithm of the minus component of the momentum. This corresponds to a fraction of the total longitudinal momentum of the target, which is the variable used in its DGLAP evolution, and also the usual physical interpretation of $\xbj$ in the parton model. In order to translate a plus momentum to a minus, one needs to have access to the correct transverse momentum scale. In the case of the whole DIS process this would naturally be $Q^2$. Thus we would want to define things in such a way that the largest evolution rapidity reached in the process is
\begin{equation}
    \label{eq:eta-rho-max}
    \eta_\text{max} \sim Y_\text{max} - \ln \frac{Q^2}{Q_0^2} = \ln \frac{1}{\xbj},
\end{equation} 
with $Y_\text{max}$ from Eq.~\eqref{eq:y-max}. Here we see that the target rapidity $\eta$ is directly related to the Bjorken-$x$ in DIS, as expected. Thus, similarly as one can think of evolution in $Y$ as evolution in $\ln W^2$, evolution in $\eta$ corresponds to evolution in $\ln 1/\xbj$.

The complication in using the target momentum fraction is that both the evolution equation and the impact factors are written  in transverse coordinate space, which is natural for the eikonal   interaction with the target. Thus the gluon transverse momentum is not very explicit in either. The usual procedure is to use an uncertainty principle argument and estimate the transverse momentum as the inverse of the corresponding transverse distance. In both the BK equation and the impact factor one integrates over transverse distances up to infinity, which would correspond to zero transverse momentum and infinite $\eta$ (for a fixed $Y$). Distances longer than some nonperturbative scale should, however, not have a significant effect on the physics. Thus we do not want large dipoles with sizes above a (soft) target transverse momentum scale $1/Q_0^2$ (the same $Q_0^2$ that we have already used) to appear in the relation between the rapidities $Y$ and $\eta$.
 In practice we are thus led to consider a dipole of size $r$ at a projectile momentum fraction corresponding to $Y$, to have a target evolution rapidity $\eta$ given by
\begin{equation} \label{eq:eta-rho-definition}
    \eta \equiv Y-\ln \frac{1}{\min\{ 1, r^2 Q_0^2 \}} = \ln \frac{W^2z_2\min\{ 1, r^2 Q_0^2 \}}{Q_0^2}.
\end{equation}
We see that with this definition we always have $\eta < Y$, which corresponds to the fact that for perturbative size dipoles $r^2 <  1/Q_0^2$ we always have less evolution in $\eta$ than in $Y$ (see \eq\eqref{eq:eta-rho-max}).

The evolution equation for the dipole amplitude in terms of the target rapidity $\eta$ was derived in Ref.~\cite{Ducloue:2019ezk} as\footnote{Compared to the recent analysis in Ref.~\cite{Ducloue:2019jmy}, we include the step function $\theta(\eta - \etaobk - \delta)$ and leave the resummation of the single transverse logarithms for future work}
\begin{multline}
    \label{eq:trbk}
     \partial_\eta \bar S(\xij{01},\eta) =
     \int \der^2 \zt  K_\text{BK}(\xij{0},\xij{1}, \xij{2}) \theta(\eta-\etaobk - \delta)
     \\
     \times
     [\bar S(\xij{02}, \eta-\delta_{02}) \bar S(\xij{21}, \eta-\delta_{21})
    %  \\
     - \bar S(\xij{01}, \eta)  ],
\end{multline}
where $\bar S$ refers to the dipole scattering matrix depending on the target rapidity $\eta$, instead of projectile rapidity $Y$. This evolution equation then needs to be provided with an initial condition at the initial rapidity $\etaobk$. We include running coupling corrections and use the kernel $K_\text{BK}(\xij{0},\xij{1}, \xij{2})$ from Eq.~\eqref{eq:bk-rc-balitsky}. The rapidity shift reads
\begin{equation}
    \label{eq:trbk-shift}
    \delta_{kl}= \max\left\{ 0, \ln \frac{\xij{01}^2}{\xij{kl}^2} \right\}.
\end{equation}
The step function with $\delta \equiv \max\{\delta_{02}, \delta_{21}\}$ ensures that the equation is a well defined initial value problem and no information about the dipole amplitude for $\eta < \etaobk$ affects the evolution. When calculating the cross section, we use  $\bar S(\rt,\eta)=\bar S(\rt,\etaobk)$ for $\eta<\etaobk$.

Equation~\eqref{eq:trbk} is the ``canonical'' BK equation from~\cite{Ducloue:2019ezk}, which contains an all-order resummation of the double collinear logarithm enhanced corrections, and thus is perturbatively correct up to an error of $\ocal(\asb^2)$. The full NLO BK evolution in target rapidity has not been solved numerically so it is not known in practice how well this resummation captures the NLO effects. In Ref.~\cite{Ducloue:2019ezk} a comparison is made between two formulations of the equation with double logarithm resummations, and the differences are minor, and mostly in the early evolution. The resummed evolution is also compared to the LO BK evolution formulated in target rapidity, and the resummed evolution is found to be notably slower.

To use the target rapidity dependent dipole amplitudes in the NLO impact factors, we simply need to replace the dipoles in the impact factor with the $\eta$-dependent dipoles. The rapidity argument is determined by using $z_2$ to obtain $Y$ with \eq\eqref{eq:y_z2}, which is then transformed into $\eta$ using~\eq\eqref{eq:eta-rho-definition}, i.e. 
\begin{equation}
\label{eq:y_eta_shift}
    S(\xij{ij}, Y) \to \bar S\left(\xij{ij}, \eta=Y-\ln \frac{1}{\min\{ 1, \xij{ij}^2 Q_0^2 \} } \right).
\end{equation}
with $Y$ defined in Eq.~\eqref{eq:y_z2}.
The regulator  ensures that the rapidity shift is always negative, consistent with the definition \eqref{eq:eta-rho-definition} and with the rapidity shift in the TBK evolution equation~\eqref{eq:trbk-shift}. 

Let us finally discuss the kinematical limits in the $z_2$ integral and their connection to the target momentum fraction probed by the process. The lower limit $z_2 > \zmin$ of the $z_2$-integral~\eqref{eq:z2min} corresponds to the lower limit of the $\eta$ values probed by the impact factor
\begin{equation}
\eta > \Yoif  + \ln \min\{ 1, r^2 Q_0^2 \}.
\end{equation}
With our choice $\Yoif=0$ the values of $\eta$ needed in the cross section always extend down to evolution rapidities before the initial condition that is imposed at $\etaobk$. In this region $\eta < \etaobk$ the dipole operators are, as in Ref.~\cite{Ducloue:2019jmy},  just frozen to the initial value: $\bar S(r,\eta < \etaobk) \equiv \bar S (r,\etaobk)$. At the upper limit $z_2=1$, on the other hand, the largest values of $\eta$ are reached for $r> 1/Q_0$, with the range in $\eta$ extending up to
\begin{equation}
    \eta < \ln \frac{W^2}{Q_0^2},
\end{equation}
which is the same as the maximum $Y$ reached in projectile momentum fraction evolution. Here let us note two things. Firstly, the $\min$ function used in defining the target momentum fraction rapidity~\eqref{eq:eta-rho-definition} prohibits $\eta$ from getting infinitely large (which would require an infinite amount of evolution) for very large dipoles $r> 1/Q_0$ that are not expected to contribute significantly to the cross section. Secondly, the amount of evolution, or largest rapidity reached, in target momentum fraction  given by $\xbj$, as in \eq\eqref{eq:eta-rho-max}, strictly speaking applies only to typical dipole sizes $r\sim 1/Q$. For larger dipoles $1/Q \lesssim r \lesssim 1/Q_0$ one actually evolves further in the target momentum fraction.

\subsection{Running coupling}
\label{sec:rc}

For the strong coupling constant in coordinate space we use the expression 
\begin{equation}
    \as(\xij{ij}^2) = \frac{4 \pi}{ \beta_0 \ln \left[ \left( \frac{\mu_0^2}{\lqcd^2}\right)^{1/c} + \left(\frac{4 C^2}{\xij{ij}^2 \lqcd^2}\right)^{1/c} \right]^c },
\end{equation}
with $\beta = (11 \nc - 2 \nf)/3$ and $\nf=3$, $\lqcd=0.241\gev$. The parameter $C^2$ controls the running coupling scale in the transverse coordinate space, i.e. $\alpha_{s}(k^2 \sim C^2/r^2)$. From Fourier analysis it has the expected value of $C^2=e^{-2 \gamma_E}$~\cite{Kovchegov:2006vj,Lappi:2012vw}. In this work, however, we take $C^2$ to be a fit parameter to absorb missing non-perturbative or higher order contributions in the modified evolution speed, similarly to previous LO fit studies~\cite{Albacete:2010sy, Lappi:2013zma}. The parameters $\mu_0$ and $c$ control how the coupling is frozen in the infrared, and we choose $\mu_0/\lqcd=2.5$ and $c=0.2$. With this choice, the coupling freezes to  $\as=0.762$ in the infrared.

We have performed fits with two different running coupling prescriptions. The first one is denoted \emph{Balitsky + smallest dipole} (Bal+SD) scheme below. In this scheme, we use the Balitsky prescription from Ref.~\cite{Balitsky:2006wa} in the BK evolution as in Eq.~\eqref{eq:bk-rc-balitsky}. In the NLO impact factor, Eq.~\eqref{eq:NLO_qg_unsub}, and in the terms resumming large transverse logarithms, Eqs.~\eqref{eq:resum-dla} and~\eqref{eq:resum-stl} in the ResumBK evolution equation, the scale is set by the smallest dipole
\begin{equation}
    \alpha_{\mathrm{s,\, sd}} \left(\xt_{01}^2, \xt_{02}^2, \xt_{21}^2 \right) = 
        \as
        \left(
            \min \left\lbrace \xt_{01}^2, \xt_{02}^2, \xt_{21}^2 \right\rbrace
        \right).
\end{equation}
Note that the Balitsky prescription reduces to the smallest dipole one when one of the dipoles is much smaller than the others.
For comparison we also use another scheme denoted as \emph{parent dipole}. Here, the  scale is always set by the size of the parent dipole, both in the evolution equation and in the impact factor.

In the LO-like $\sigmaltdip$ term of the impact factor, Eq.~\eqref{eq:NLO_dip}, there are no daughter dipoles in the scattering state. For this term the smallest dipole scheme is equivalent with the parent dipole scheme.

\subsection{Initial conditions}
\label{sec:ic}
The initial condition for the (projectile momentum fraction) BK evolution is parametrized at rapidity $Y=\Yobk$. We use the MV$^\gamma$ parametrization used previously in similar fits \cite{Albacete:2009fh, Albacete:2010sy}, and write the initial condition as
\begin{multline}
    \label{eq:bk-ic}
    S(\xij{ij}, Y = \Yobk) =
    \\
    \exp 
        \left[ 1
            - \frac{\left(\xij{ij}^2Q_{s,0}^2\right)^\gamma}{4} \right.
            % \\
            \left. \ln \left( \frac{1}{|\xij{ij}| \Lambda_\text{QCD}}
             +  e   \right)   \right]. 
\end{multline}
The fit parameters in the initial condition are $\qso^2$ which controls the saturation scale at the initial $x$, and the anomalous dimension $\gamma$ which determines the shape of the dipole amplitude at small $|\xij{ij}|$. We note that this parametrization results in both a negative unintegrated gluon distribution and negative particle production cross sections in proton-nucleus collisions at high transverse momenta if $\gamma>1$. As the inclusive DIS measurements are not sensitive to asymptotically small dipoles, we do not consider our dipole amplitude to be valid in that region and as such, having an anomalous dimension $\gamma>1$ is acceptable. The practical interpretation of $\gamma$ in our fit is that it controls the shape of the dipole amplitude in the transient region $r\sim 1/Q_s$. The leading order BK fits to HERA data generally prefer $\gamma \sim 1.1$~\cite{Albacete:2010sy,Lappi:2013zma}, and similar results were found in recent fits where the BK equation with some higher order corrections resummed~\cite{Ducloue:2019jmy} was used. For a detailed discussion related to the Fourier positivity of the dipole amplitude, the reader is referred to Ref.~\cite{Giraud:2016lgg}.
\label{sec:positivity}

For the local resummed projectile momentum fraction (ResumBK) evolution, the resummation should also  in principle affect  the initial condition~\cite{Iancu:2015vea}. However, as the initial condition is in any case a non-perturbative input, we will use the same parametrization of Eq.~\eqref{eq:bk-ic} also for solving the ResumBK equation.

For target momentum fraction evolution, the initial condition for the evolution \eqref{eq:trbk} corresponds to the scattering amplitude  $\tilde S(r,\eta=\etaobk)$ at some rapidity $\etaobk$. 
We use the same parametrization, \eq\eqref{eq:bk-ic}, as in the case of projectile rapidity evolution.

\section{Available DIS data}
\label{sec:data}
The HERA experiments H1 and ZEUS have published their combined measurements for the reduced cross section $\sigma_r$ in Refs.~\cite{Aaron:2009aa,Abramowicz:2015mha}. Additionally, the charm and bottom quark contributions to the fully inclusive data are available~\cite{Abramowicz:1900rp,H1:2018flt}. As the impact factors at next to leading order accuracy in the massive quark case are not available, we only calculate the light quark contribution to the photon-proton cross section. In the leading order fits~\cite{Albacete:2010sy,Lappi:2013zma} it has been possible to obtain a good description of the fully inclusive data with only light quarks, even though the charm contribution is significant (parametrically up to $\sim 40\%$ at $Q^2\gg m_c^2$). On the other hand, the leading order fits aiming to simultaneously describe the total and charm structure function data require separate parameters (e.g. different  transverse areas)  for the light and charm quarks~\cite{Albacete:2010sy}, or an additional effective soft and non-perturbative  contribution~\cite{Berger:2011ew,Mantysaari:2018zdd}. 

In this work we consider two different setups. First, we follow the strategy that has been successfully used at leading order and calculate the light quark contribution to the structure functions, and compare with the inclusive HERA data from Ref.~\cite{Aaron:2009aa}. We note that the newer combined dataset containing data from the HERA-II run is also available~\cite{Abramowicz:2015mha}, but at low $x$ and moderate $Q^2$ the two datasets result in very similar fits (see e.g. Ref.~\cite{Mantysaari:2018nng}).

As a second approach, we construct an interpolated dataset that only contains the light quark contribution. Since the charm and bottom data are not measured in the same kinematical $x,Q^2$ bins as the inclusive data, it is not possible to just subtract the heavy quark contribution from the fully inclusive cross section. Instead, we use a leading order dipole model fit from Ref.~\cite{Mantysaari:2018nng}, where the Bjorken-$x$ and dipole size $\rt$ dependence is described using the so called IPsat parametrization~\cite{Kowalski:2003hm}. This parametrization includes a smooth matching to the DGLAP evolution~\cite{Gribov:1972ri,Gribov:1972rt,Altarelli:1977zs,Dokshitzer:1977sg} in the dilute region, and at large dipoles or densities the scattering amplitude saturates to unity. The advantage of this parametrization is that it results in an excellent description of both inclusive and heavy quark datasets. Consequently, it can be used to interpolate the charm and bottom contributions to the structure functions.  We use this parametrization to subtract the heavy quark contributions from the measured reduced cross section.    We then use this interpolated  light quark--only  data in the NLO fits. 
In our procedure we do not modify the uncertainties of the inclusive data in the subtraction (another possible  approach would be to reduce the uncertainties proportionally). This is not really a consistent treatment for the errors; ultimately only the experimental collaborations would be in a position to correctly take into account the correlation between errors in the total and heavy quark data.  Thus the errors and consequently $\chi^2$ values in the light-quark fits are not correct statistically.  However, we expect the magnitude of the uncertainties to only affect the final fit qualities, and to have only a limited effect on the extracted best fit parameter values and the interpretation in terms of physics. This detail must be kept in mind for the interpretation of the $\chi^2$ values from the light quark fits.

\begin{figure}[tb]
% \centering
        \includegraphics[width=0.48\textwidth]{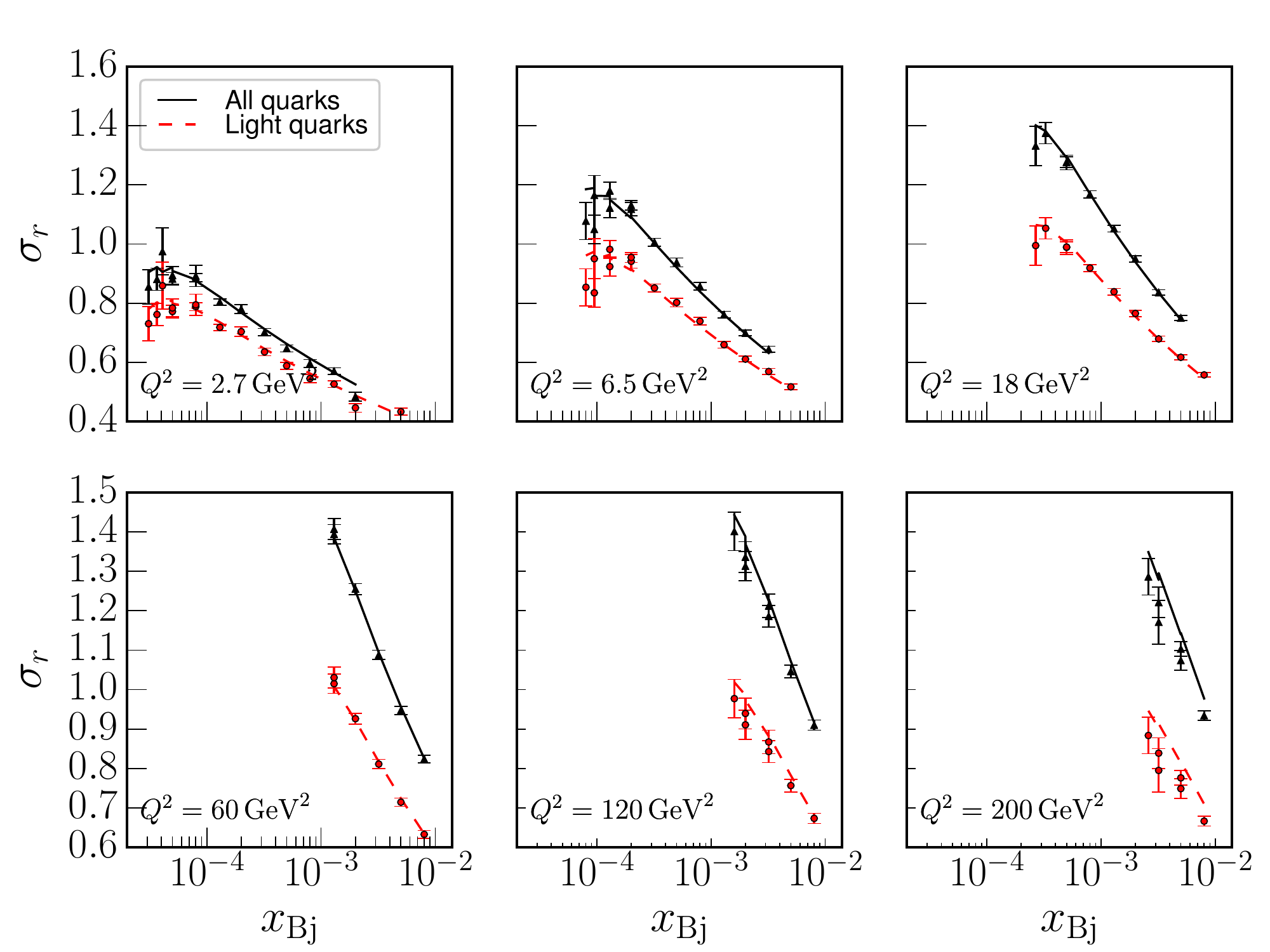}
                \caption{Total reduced cross section (black triangles) from Ref.~\cite{Aaron:2009aa} and interpolated light quark pseudodata (red circles) in a few $Q^2$ bins. The solid and dashed lines show the calculated cross sections from the IPsat fit that are used to generate the pseudodata.  }
        \label{fig:light_pseudodata}
\end{figure}

The total reduced cross section for some $Q^2$ bins from HERA~\cite{Aaron:2009aa} is shown in Fig.~\ref{fig:light_pseudodata}, and compared with the result obtained by the IPsat fit mentioned above. The description of the data is excellent. The interpolated light quark data in the same kinematics is also shown, and compared to the light quark reduced cross section computed using the same IPsat fit.

When fitting the initial condition for the BK evolution, we consider datapoints in the region $0.75 < Q^2 < 50 \gev^2$ at $x<0.01$. This results in $N = 187$ datapoints to be included in the fit. Although the correlation matrix for the experimental uncertainties is available~\cite{Aaron:2009aa}, we do not take these correlations into account as we expect it to have only a negligible effect in our fits.

\section{Fit results}
\label{sec:fitresults}

In this section we will look at our fit results. The discussion is divided first by the data that is being fitted, followed by a comparison of the evolution prescriptions in the kinematical domain accessible in future DIS experiments, which lies outside the HERA region included in the fits.

Let us first recall the essential details of our fit schemes. The choice of a fit scheme consists of  the version of the BK evolution equation (discussed in Secs.~\ref{sec:kcbk}, \ref{sec:resum} and \ref{sec:trbk}),
the running coupling scheme (see Sec.~\ref{sec:rc}), and the starting point of the BK evolution, parametrized in terms of $\Yobk$ or $\etaobk$. The fit results in values for the free parameters characterizing the initial condition as discussed in Sec.~\ref{sec:ic}: $\qso^2$, $\sigma_0$ and $\gamma$, and in a value for the parameter $C^2$ in the scale of the running coupling, see Sec.~\ref{sec:rc}.

Our main fit results are presented in Tables~\ref{tab:fits-kcbk}, \ref{tab:fits-resumbk} and~\ref{tab:fits-trbk} classified by the BK equation used, with secondary and tertiary grouping keys being the running coupling scheme and $\Yobk$ (or $\etaobk$) controlling the rapidity scale of the BK initial condition  used in the fits. The saturation scale $Q_s^2$ defined  as $N(r^2=2/Q_s^2)=1-e^{-1/2}$ is also shown  at fixed projectile rapidity $Y=\ln \frac{1}{0.01}$. We will first discuss in the next subsection the fits to the full HERA reduced cross section data, and in the following subsection the fits to the interpolated light quark pseudodata presented in Sec.~\ref{sec:data} and labeled as \emph{light-q} in the tables where the fit results are shown. The two datasets differ enough to to warrant their own discussion.

\begin{table*}[ht]
    \begin{tabular}{ |L{\mycol}|L{1.1\mycol}|R{\mycol}||R{\mycol}|R{\mycol}|R{\mycol}|R{\mycol}|R{\mycol}||R{1.8\mycol}| }
        \hline
        Data & $\alpha_s$ & \multicolumn{1}{l||}{$\Yobk$} & \multicolumn{1}{l|}{$\chi^2/N$} & \multicolumn{1}{l|}{$Q_{s,0}^2$} & \multicolumn{1}{l|}{$C^2$} & \multicolumn{1}{l|}{$\gamma$} & \multicolumn{1}{l||}{$\sigma_0/2$} & \multicolumn{1}{l|}{$Q_s^2(Y=\ln \frac{1}{0.01})$}\\
         &  &  &  & \multicolumn{1}{l|}{$[\mathrm{GeV}^2]$} & & & \multicolumn{1}{l||}{[mb]} & \multicolumn{1}{l|}{$[\mathrm{GeV}^2]$}\\ \hline
        HERA    & parent & $\ln \frac{1}{0.01}$ & 1.85 & 0.0833 & 3.49 & 0.98  & 9.74   & 0.11 \\ %\hline
        light-q & parent & $\ln \frac{1}{0.01}$ & 1.58 & 0.0753 & 37.7 & 1.25  & 18.41  & 0.11 \\ %\hline
        HERA    & parent & 0                  & 1.24 & 0.0680 & 79.9 & 1.21  & 18.39    & 0.20 \\ %\hline
        light-q & parent & 0                  & 1.18 & 0.0664 & 1340 & 1.47  & 27.12    & 0.14 \\ %\hline
        \hline
        HERA    & Bal + SD & $\ln \frac{1}{0.01}$ & 1.89 & 0.0905 & 0.846   & 1.21  & 8.68 & 0.13 \\ %\hline
        light-q & Bal + SD & $\ln \frac{1}{0.01}$ & 2.63 & 0.0720 & 1.91    & 1.55  & 12.44  & 0.11 \\ %\hline
        HERA    & Bal + SD & 0                    & 1.49 & 0.1114 & 0.846   & 1.94  & 8.53   & 0.26 \\ %\hline
        light-q & Bal + SD & 0                    & 1.69 & 0.1040 & 2.87    & 7.70  & 12.09  & 0.14 \\ \hline
    \end{tabular}
    \caption{Fits to HERA and light quark data with the Kinematically Constrained BK evolution (KCBK). }
    \label{tab:fits-kcbk}
\end{table*}

\begin{table*}[ht]
    \begin{tabular}{ |L{\mycol}|L{1.1\mycol}|R{\mycol}||R{\mycol}|R{\mycol}|R{\mycol}|R{\mycol}|R{\mycol}||R{1.8\mycol}| }
        \hline
        Data & $\alpha_s$ & \multicolumn{1}{l||}{$\Yobk$} & \multicolumn{1}{l|}{$\chi^2/N$} & \multicolumn{1}{l|}{$Q_{s,0}^2$} & \multicolumn{1}{l|}{$C^2$} & \multicolumn{1}{l|}{$\gamma$} & \multicolumn{1}{l||}{$\sigma_0/2$} & \multicolumn{1}{l|}{$Q_s^2(Y=\ln \frac{1}{0.01})$}\\
         &  &  &  & \multicolumn{1}{l|}{$[\mathrm{GeV}^2]$} & & & \multicolumn{1}{l||}{[mb]} & \multicolumn{1}{l|}{$[\mathrm{GeV}^2]$}\\ \hline
        HERA    & parent & $\ln \frac{1}{0.01}$ & 2.24 & 0.0964 & 1.21 & 0.98  & 7.66   & 0.13 \\ %\hline
        light-q & parent & $\ln \frac{1}{0.01}$ & 1.62 & 0.0755 & 11.7 & 1.24  & 16.53  & 0.11 \\ %\hline
        HERA    & parent & 0    & 1.12 & 0.0721 & 89.5 & 1.37  & 19.68                  & 0.21 \\ %\hline
        light-q & parent & 0    & 1.18 & 0.0794 & 1480 & 1.92  & 26.69                  & 0.18 \\ %\hline
        \hline
        HERA    & Bal + SD & $\ln \frac{1}{0.01}$ & 2.37 & 0.0950 & 0.313 & 1.24 & 7.85  & 0.14 \\ %\hline
        light-q & Bal + SD & $\ln \frac{1}{0.01}$ & 2.21 & 0.0796 & 0.684 & 1.81 & 11.34 & 0.13 \\ %\hline
        HERA    & Bal + SD & 0    & 2.35 & 0.0530 & 0.486 & 1.56 & 10.10                 & 0.23 \\ %\hline
        light-q & Bal + SD & 0    & 3.19 & 0.0566 & 1.27  & 9.35 & 14.27                 & 0.13 \\ \hline
    \end{tabular}
    \caption{Fits to HERA and light quark data with local projectile momentum fraction evolution (ResumBK).}
    \label{tab:fits-resumbk}
\end{table*}

\begin{table*}[ht]
    \begin{tabular}{ |L{\mycol}|L{1.1\mycol}|R{\mycol}||R{\mycol}|R{\mycol}|R{\mycol}|R{\mycol}|R{\mycol}||R{1.8\mycol}| }
        \hline
        Data & $\alpha_s$ & \multicolumn{1}{l||}{$\etaobk$} & \multicolumn{1}{l|}{$\chi^2/N$} & \multicolumn{1}{l|}{$Q_{s,0}^2$} & \multicolumn{1}{l|}{$C^2$} & \multicolumn{1}{l|}{$\gamma$} & \multicolumn{1}{l||}{$\sigma_0/2$} & \multicolumn{1}{l|}{$Q_s^2(Y=\ln \frac{1}{0.01})$}\\
         &  &  &  & \multicolumn{1}{l|}{$[\mathrm{GeV}^2]$} & & & \multicolumn{1}{l||}{[mb]} & \multicolumn{1}{l|}{$[\mathrm{GeV}^2]$}\\ \hline
        HERA    & parent & $\ln \frac{1}{0.01}$ & 2.76 & 0.0917  & 0.641 & 0.90 & 6.19  & 0.11 \\ %\hline
        light-q & parent & $\ln \frac{1}{0.01}$ & 1.61 & 0.0729  & 14.4 & 1.19 & 16.45  & 0.10 \\ %\hline
        HERA    & parent & 0    & 1.03 & 0.0820  & 209  & 1.44 & 19.78                  & 0.23 \\ %\hline
        light-q & parent & 0    & 1.26 & 0.0731  & 8050  & 1.86 & 29.84                 & 0.16
        \\ \hline
        HERA    & Bal + SD & $\ln \frac{1}{0.01}$ & 2.48 & 0.0678 & 1.23 & 1.13 & 10.43  & 0.09 \\ %\hline
        light-q & Bal + SD & $\ln \frac{1}{0.01}$ & 1.90 & 0.0537 & 3.55  & 1.59 & 16.85 & 0.08 \\ %\hline
        HERA    & Bal + SD & 0    & 2.77 & 0.0645 & 3.67 & 6.37 & 14.14                  & 0.15 \\ %\hline
        light-q & Bal + SD & 0    & 1.82 & 0.0690 & 822  & 8.35 & 29.26                  & 0.14 \\ \hline
    \end{tabular}
    \caption{Fits to HERA and light quark data with Target momentum fraction BK (TBK) evolution. Note that the saturation scale $Q_s^2$ is extracted at fixed projectile rapidity $Y$ to allow comparisons with the projectile momentum fraction evolutions.}
    \label{tab:fits-trbk}
\end{table*}

\subsection{Fitting the HERA reduced cross section}

\begin{figure}[tb]
% \centering
        \includegraphics[width=0.48\textwidth]{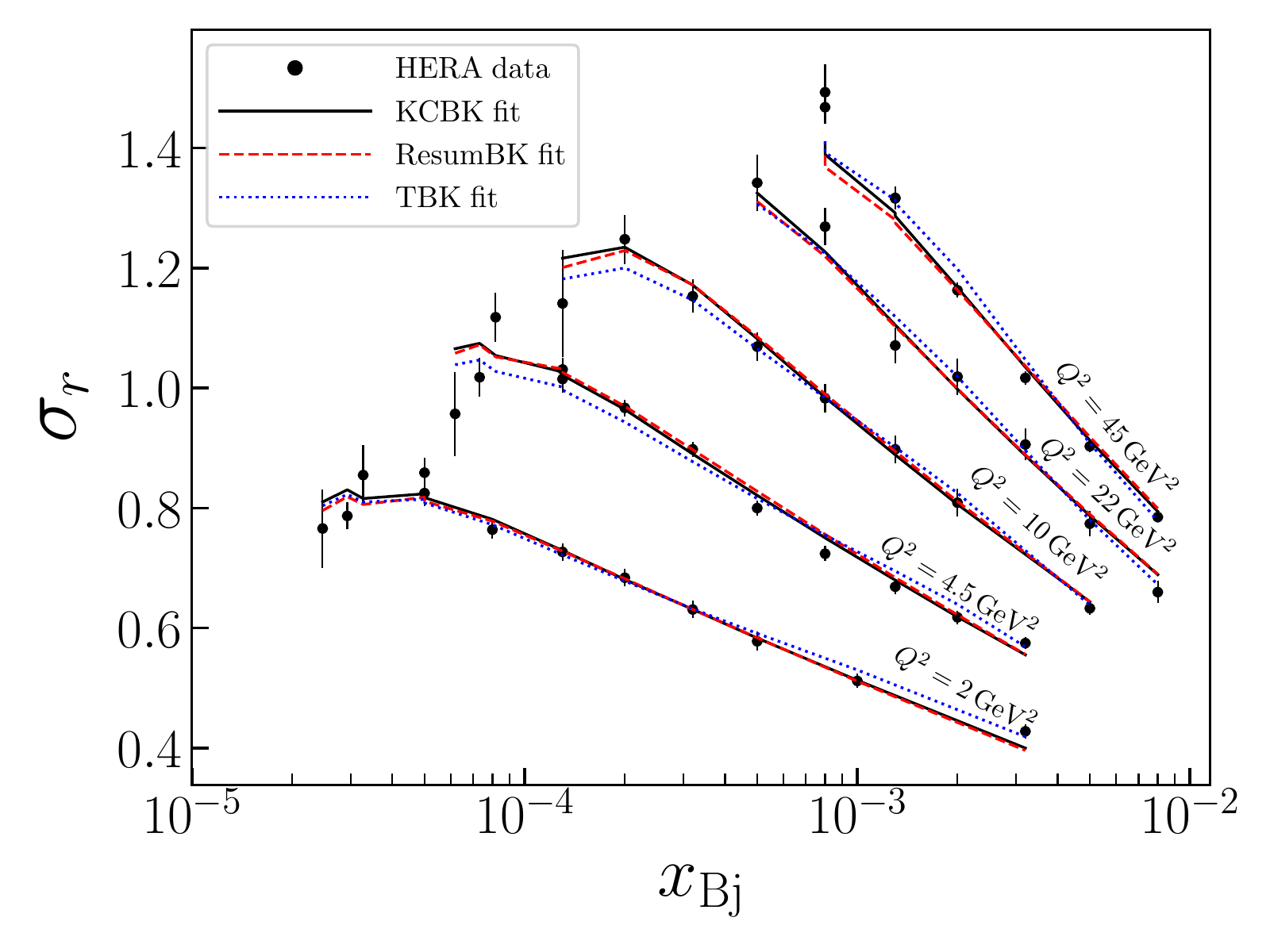}
                \caption{Reduced cross section obtained using the fits with different BK evolutions compared with the HERA data~\cite{Aaron:2009aa}. Balitsky + smallest dipole running coupling is used, with $\Yobk= \ln1/0.01$.}
        \label{fig:sigmar-hera-x001-pd}
\end{figure}

Before we discuss the results and their systematic features in more detail we show in \fig\ref{fig:sigmar-hera-x001-pd} that all three BK evolutions combined with next to leading order impact factors are capable of describing the HERA data equally well. The results shown are obtained using the Bal+SD running coupling, and $\Yobk=\etaobk= \ln1/0.01$, but excellent fit results are obtained with other scheme choices, too. Even though the resulting parametrizations for the dipole at initial rapidity can differ significantly, the resulting reduced cross sections are mostly indistinguishable.

We first present in Table~\ref{tab:fits-kcbk}  the fit results obtained using the kinematically constrained BK equation as discussed in Sec.~\ref{sec:kcbk}. We find a very good description ($\chi^2/N = 1.49$) of the HERA data using our main setup with the Bal+SD prescription and $\Yobk = 0$. We consider this as our preferred HERA data fit, with a BK equation derived  in the same framework as the impact factor, a theoretically preferred running coupling scheme, and only one starting scale $\Yoif=\Yobk=0$. We note that starting the BK evolution at $\Yobk=\ln\frac{1}{0.01}$ (and freezing the dipole at smaller rapidities) results in an equally good fit. This suggests that we are only weakly sensitive to the details of extrapolation scheme used to describe the dipole amplitude in the region $\Yoif < Y < \Yobk$.  The parameter $C^2$ controlling the evolution speed is not required to be large as it is in the case of leading order fits, where one generally finds $C^2\sim 10$~\cite{Albacete:2010sy,Lappi:2013zma}. Instead, we find $C^2\approx 0.85$, which is of the same ballpark as the general estimate $C^2=e^{-2\gamma_E}\approx 0.3$~\cite{Kovchegov:2006vj,Lappi:2012vw}.

As seen in Table.~\ref{tab:fits-kcbk}, larger values of $C^2$ are required in the parent dipole scheme fits. This is expected, as $C^2$ maps the coordinate space scale $\xij{ij}^2$ to momentum space $C^2/\xij{ij}^2$, and in the parent dipole scheme the coordinate space scale is generically larger. Consequently a larger $C^2$ is needed to render the strong coupling values, and the resulting evolution speeds, comparable between the coupling constant scheme choices.

    \begin{figure}[tb]
    \centering
            \includegraphics[width=0.48\textwidth]{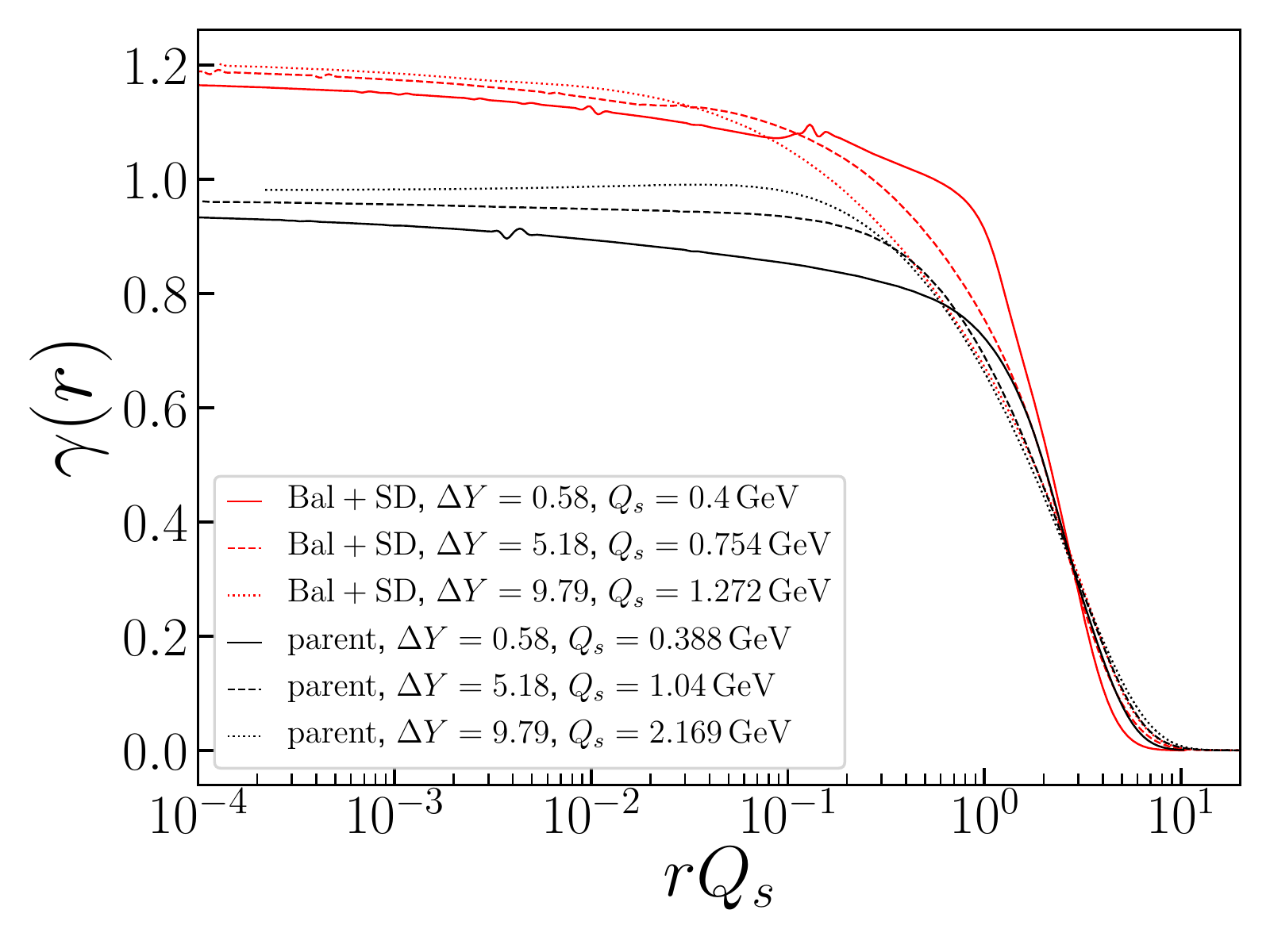}
                    \caption{
                    Anomalous dimension evolution with KCBK, $\Yobk = \ln 1/0.01$. Computed from fits to HERA data. 
                    }
            \label{fig:anomdim-kcbk-x001}
    \end{figure}

We generically find $\gamma>1$ at the initial condition, with the exception $\gamma\approx 1$ found in the case where the evolution starts  at $\Yobk=\ln \frac{1}{0.01}$ and the parent dipole prescription for the running coupling is used. We note that $\gamma>1$ is also required in the leading order fits to obtain a $Q^2$ dependence at the initial condition compatible with the HERA data~\cite{Albacete:2010sy,Lappi:2013zma}.
The disadvantage of an initial condition with $\gamma>1$ is that, as discussed in Sec.~\ref{sec:positivity}, it results in the unintegrated gluon distribution not being positive definite at large transverse momenta.

To understand why different running coupling prescriptions result in  different initial anomalous dimensions, we  study the slope of the dipole defined as
\begin{equation}
    \gamma(r) = \frac{\der \ln N(r)}{\der \ln r^2}.
\end{equation}
For the KCBK fits this is shown in Fig.~\ref{fig:anomdim-kcbk-x001} as a function of dimensionless dipole size $rQ_s$.
The kinematically constrained BK equation is found to keep the anomalous dimension (slope at small $r$) approximatively constant at very small $r$, unlike the leading order BK equation. A similar effect was found in case of the ResumBK equation in Ref.~\cite{Lappi:2016fmu}.

At  intermediate  $r\sim 1/Q_s$ which dominates the cross section, there is clear evolution towards an asymptotic shape. Let us first focus on results where the smallest dipole Bal+SD coupling is used. Here, the anomalous dimension is large at the initial condition and the evolution decreases the slope at intermediate $rQ_s$, which results in the cross section growing more rapidly with $Q^2$. If the BK evolution is started at the rapidity scale from which there is a long evolution before entering the data region (i.e. $\Yobk=0$), a larger initial anomalous dimension is required in order to obtain the shape dictated by the $Q^2$ dependence of the HERA structure function data around $r\sim 1/Q_s$.

Let us then consider the evolution with the parent dipole prescription. In this case, we start from a relatively small $\gamma=0.98$, and the evolution increases the slope  at small (but not asymptotically small) $r$. This can be seen to stem from the fact that in the parent dipole prescription the  coupling, and consequently the evolution speed of the dipole amplitude $N(r)$, grows more as a function of parent dipole size $r$ in comparison to other running coupling prescriptions. 
At larger $r$, the slope evolves only slightly.  After a few units of rapidity evolution, the dipole amplitudes have the same shape in the $r\sim 1/Q_s$ region independently of the running coupling prescription. This is expected, as $r\sim  1/Q_s$ size dipoles dominate when calculating the structure functions in HERA kinematics.

Next we move to the local projectile momentum fraction (ResumBK) fits, the results of which are shown in Table~\ref{tab:fits-resumbk}. In general, the results are close to the ones previously discussed in case of the kinematically constrained BK equation. This is not surprising, as both equations are designed to include the same subset of higher order corrections enhanced by large double transverse logarithms. Similarly to the preferred fit with KCBK, in the ResumBK fit to HERA data with Bal+SD and $\Yobk = 0$ the obtained $C^2$ is quite small at $C^2\approx 0.49 \sim e^{-2 \gamma_E}$ and the anomalous dimension is large, $\gamma=1.56$. The obtained anomalous dimension values behave similarly as in the case of KCBK, and there similarly seems to be a systematic preference for smaller $\sigma_0/2$ with the Balitsky + smallest dipole coupling.

The ResumBK equation evolves generically more slowly than the KCBK equation, which is reflected in  the required $C^2$ values being smaller (except in the case $\Yobk=0$ with parent dipole prescription when the $C^2$ values are comparable). This is a consequence of  the ResumBK equation including an additional resummation of some single transverse logarithms. The main effect of this resummation is that it results in a slower evolution, see discussion in Sec.~\ref{sec:resum}. We have confirmed numerically that if the resummation of single transverse logarithms is not included, our fit results are almost intact, except that a larger value for the parameter $C^2$ is obtained.

In both ResumBK and KCBK fits with Bal+SD running coupling, the obtained values for the proton transverse area $\sigma_0/2$ are generally smaller than what is found in leading order fits with similar running coupling schemes, with or  without a resummation of large transverse logarithms~\cite{Albacete:2010sy,Lappi:2013zma,Ducloue:2019jmy}. The obtained saturation scales at $Y=\ln \frac{1}{0.01}$, on the other hand, are comparable to the leading order fit results. In the LO fits, one typically obtains $\sigma_0/2\sim 16 $ mb (proton sizes comparable to our results were found in the leading order fit presented in Ref.~\cite{Albacete:2015xza} where double logarithmic corrections were resummed in the BK equation similarly as in our setup). 

We note that the proton transverse area can in principle be obtained by studying the squared momentum transfer $t$ dependence of exclusive vector meson production. If the cross section is written as $e^{-B_D |t|}$ at small $|t|$, the HERA measurements on $\mathrm{J}/\psi$ production~\cite{Chekanov:2004mw,Alexa:2013xxa} give $B_D\approx 4 \gev^{-2}$. Depending on the assumed proton density profile, this corresponds to $\sigma_0/2 \approx 9.8 \dots 19.6$~mb (using Gaussian or a step function profile). As the vector meson $t$ spectra are not measured precisely enough especially at large $|t|$, the exact form of the proton density profile can not be deduced. Consequently, we find that all obtained values for the proton transverse size $\sigma_0/2$ in our fits to HERA reduced cross section data are compatible with the $\mathrm{J}/\psi$ spectra. However, we also note that the step function profile is not really favored by the HERA data~\cite{Kowalski:2006hc}. Thus one would prefer values that are in the lower part of the range $\sigma_0/2 \approx 9.8 \dots 19.6$~mb. Indeed,  especially with the Balitsky + smallest dipole running coupling, our fit results for the proton size also favor such smaller target sizes for the proton.

As we are neglecting the impact parameter dependence, we can not compute the evolution of the proton transverse area and consequently use a fixed $\sigma_0/2$ at all $\xbj$. We note that the HERA vector meson production data~\cite{Alexa:2013xxa,Chekanov:2002xi}  suggest that the transverse area depends logarithmically on the center-of-mass energy. This growth is effectively included in the energy dependence of the proton saturation scale in our framework.

Let us finally discuss the results obtained with the third evolution equation considered in this work, the BK equation formulated in terms of the target momentum fraction (TBK). 
The fit results in this case are shown in Tab.~\ref{tab:fits-trbk}. While the fit qualities overall are quite similar to the projectile momentum fraction setups, we find some  departures from the shared qualitative features of the KCBK and ResumBK fits. With the Balitsky + smallest dipole running coupling the TBK evolution needs to be slowed down more with a larger values of $C^2$ compared to KCBK and ResumBK equations. The TBK fit with parent dipole coupling is more mixed in this respect: with $\etaobk = 0$ setups the $C^2$ values are quite a bit larger but then with $\etaobk =\ln \frac{1}{0.01}$ the HERA data fit is found to require only a small $C^2$.

Comparing the initial conditions with $\etaobk = 0$ and $\etaobk = \ln \frac{1}{0.01}$ we see that every evolution starts from a significantly larger anomalous dimension when $\etaobk = 0$. This difference is more pronounced compared to the previously studied KCBK and ResumBK equations.
This is because the TBK evolution drives the dipole towards the asymptotic shape with a small anomalous dimension $\gamma \sim 0.6$~\cite{Ducloue:2019ezk}. This behavior is similar to the leading order BK equation, in which case it is already known that the asymptotic shape can not be used to parametrize the initial condition~\cite{Albacete:2010sy}\footnote{In \cite{Kuokkanen:2011je} an asymptotic form of the initial condition produces working results only when a significant additional ``energy conservation'' correction in the BK evolution is used.}. Indeed the development of the geometric scaling regime independently of the initial condition is a theoretically  attractive feature of the TBK formulation. However, HERA data seems to prefer to lie in the preasymptotic regime in the fits. Thus, especially  in the fits with more evolution before the data region (smaller $\etaobk$), one needs to slow down the evolution more and start with a significantly larger anomalous dimension in order to still have a transient form of the dipole amplitude in the data regime.

\begin{figure}[tb]
    % \centering
            \includegraphics[width=0.48\textwidth]{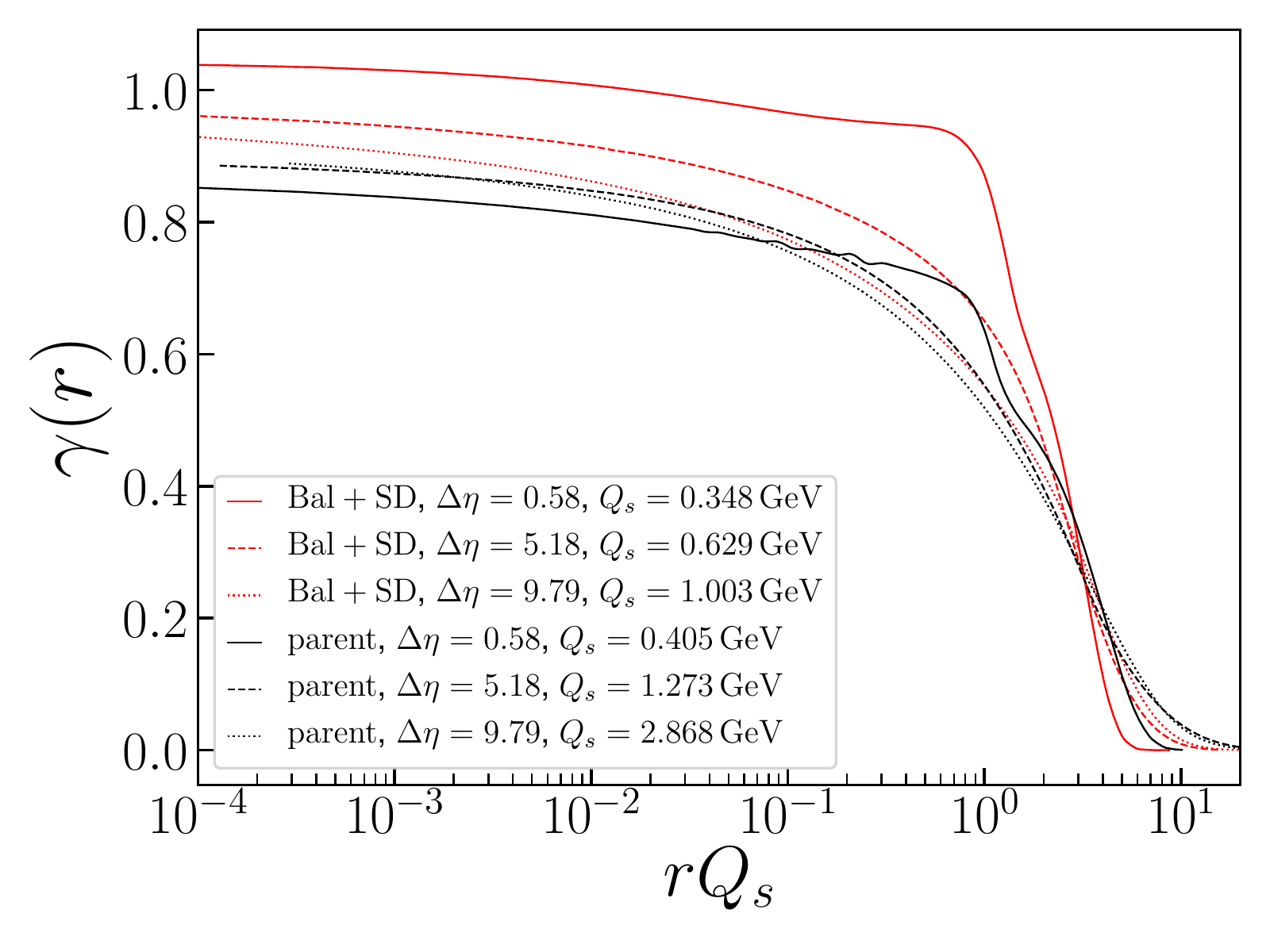}
                    \caption{
                    Anomalous dimension evolution with TBK, using the initial conditions parametrized at $\etaobk = \ln\frac{1}{0.01}$. The evolved rapidity range from the initial condition is denoted by $\Delta \eta$.
                    }
            \label{fig:anomdim-trbk-x001}
    \end{figure}

The evolution of the dipole slope in TBK evolution is shown in Fig.~\ref{fig:anomdim-trbk-x001} at different evolution rapidities $\eta$. Unlike in the case of KCBK equation discussed earlier and shown in Fig.~\ref{fig:anomdim-kcbk-x001}, the slope of the dipole from the TBK evolution is decreasing with both running couplings in the $r \sim 1/Q_s$ regime.  As shown in Ref.~\cite{Ducloue:2019ezk}, the asymptotic anomalous dimension $\gamma \sim 0.6$ is obtained only at very large rapidities, and at least the Bal+SD coupling case can be seen to be evolving towards this asymptotic value.  In the rapidity range relevant in HERA or even LHeC kinematics the asymptotic anomalous dimension is not reached. With the parent dipole coupling, a significantly longer evolution is needed before evolution towards the asymptotic shape at small $r$ becomes visible in the best fit case with a small anomalous dimension $\gamma=0.9$ in the initial condition.

    \begin{figure}[tb]
    % \centering
            \includegraphics[width=0.48\textwidth]{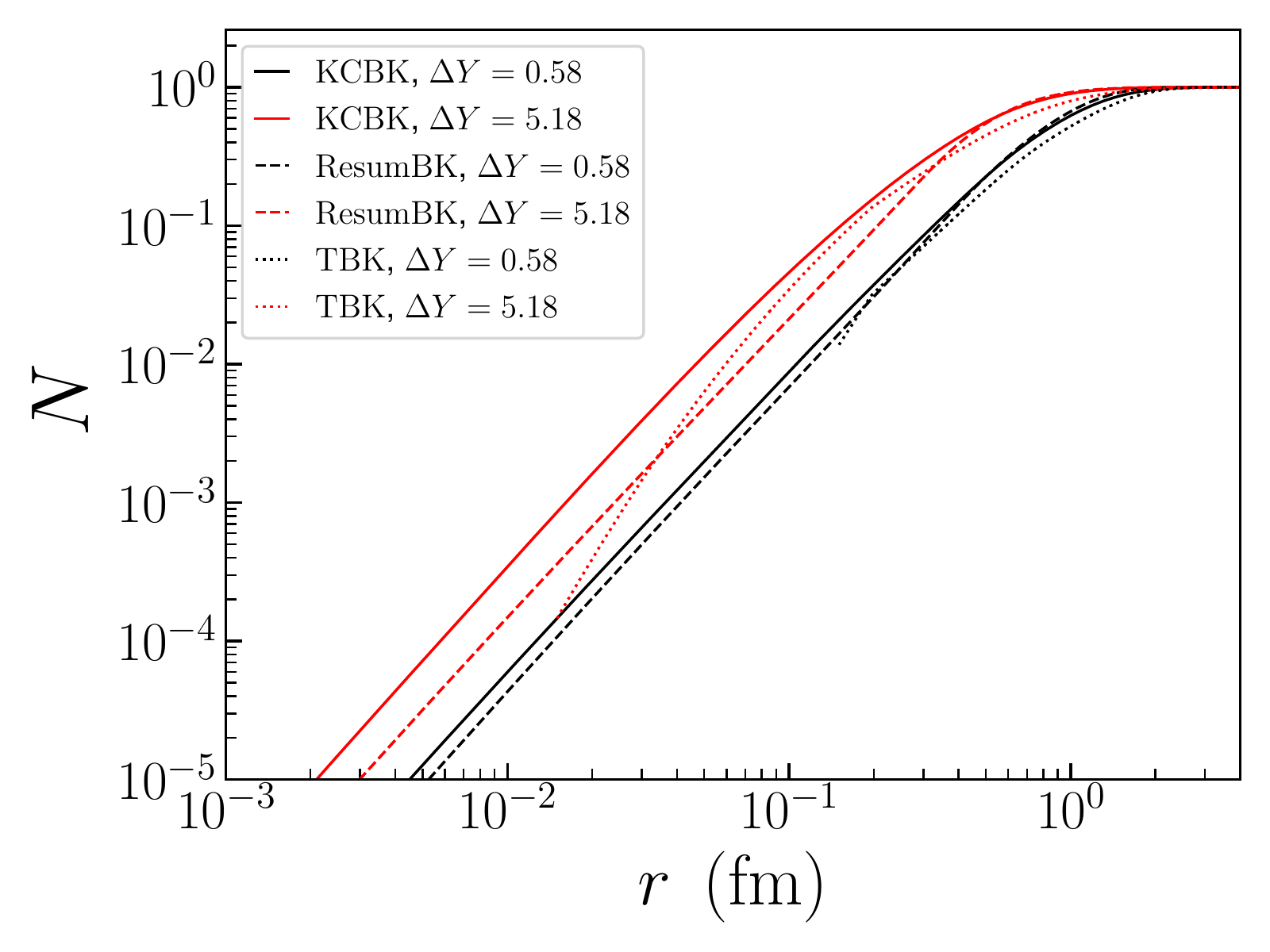}
                    \caption{
                    Dipole amplitudes of the three BK equations at an early and later stage in the evolution at constant evolution rapidities $Y = \Yobk + \Delta Y$, with  TBK solutions in $\eta$ shifted into $Y$. Balitsky + smallest dipole running coupling is used, with the initial conditions from the fits with $\Yobk = \etaobk = \ln \frac{1}{0.01}$.
                    }
            \label{fig:amplitude-hera-x001}
    \end{figure}

The dipole amplitudes at different evolution rapidities as a function of dipole size are shown in Fig.~\ref{fig:amplitude-hera-x001}. Here, results obtained using the all three considered evolution equations are shown at fixed projectile rapidity $Y=\Yobk + \Delta Y$. The solution to the TBK evolution is shifted from the target rapidity $\eta$ to the projectile rapidity $Y$ by performing the shift~\eqref{eq:y_eta_shift}. The shifted TBK solutions are shown in the region where $\eta > \etaobk$. In the region where the dipole amplitude is not small, all evolution equations result in comparable dipole amplitudes. This is expected, as all the shown dipoles result in a compatible description of the HERA structure function data. 

At small dipole sizes that do not significantly contribute to the structure functions some differences appear. Despite the fact that KCBK and ResumBK equations have very similar initial conditions the resulting amplitudes differ significantly for small dipoles. This is mostly driven by the resummation of the single transverse logarithms not included in the kinematically constrained BK equation, as this resummation is more important at small parent dipole size $r$. At very small dipoles the TBK evolved dipole also differs significantly from the other dipoles when the shift from target rapidity, Eq.~\eqref{eq:trbk-shift}, results in the dipole being evaluated close to the initial condition. If the parent dipole scheme for the running coupling were used, the differences between the dipoles obtained from the different evolution equations would be significantly reduced, as in that scheme the coupling constant is generically smaller at small $r$ and differences between the evolution equations are suppressed by the small $\as$.

\subsection{Fitting the interpolated light quark reduced cross section}

\begin{figure}[tb]
% \centering
        \includegraphics[width=0.48\textwidth]{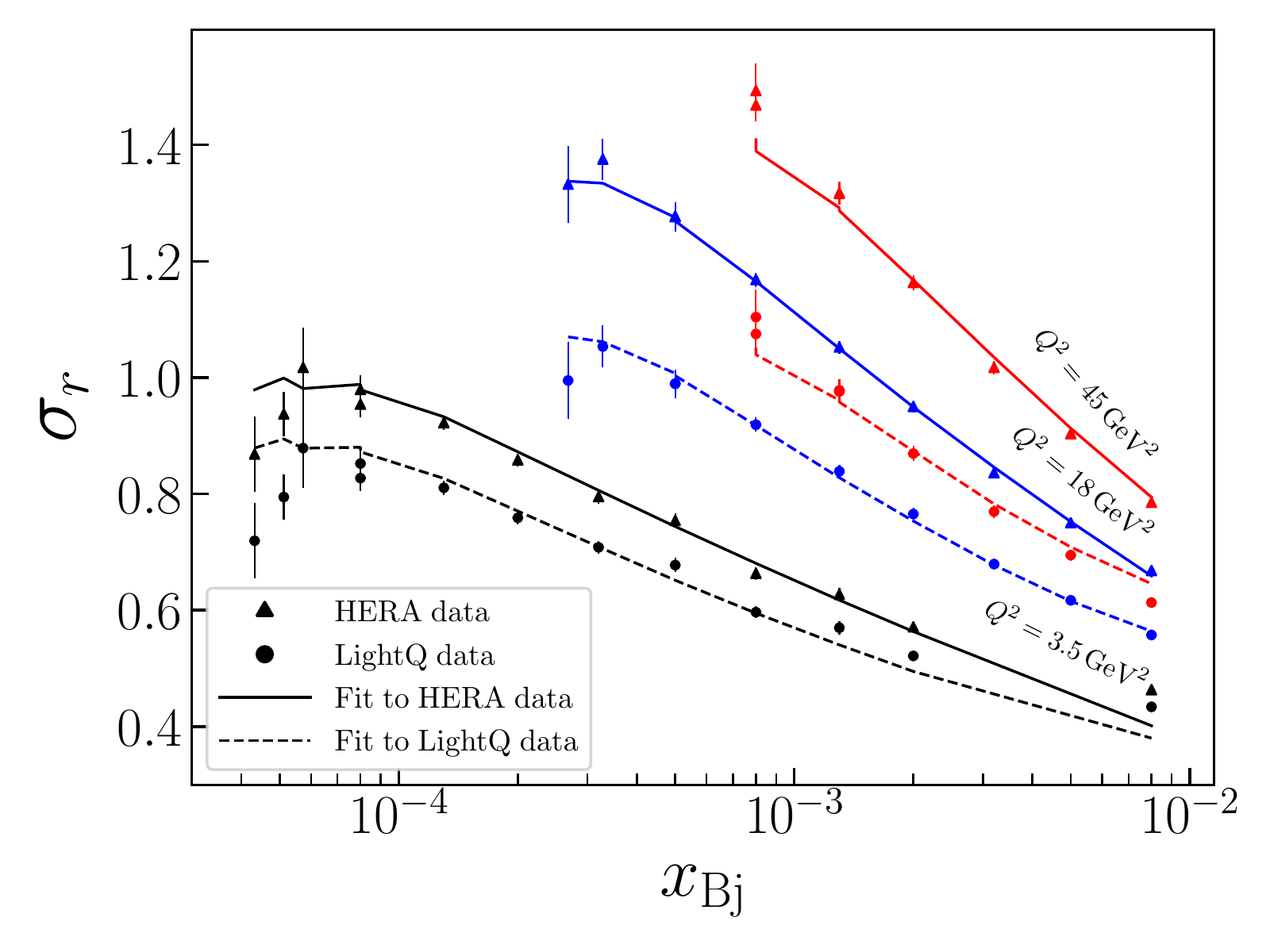}
                \caption{Total and light quark reduced cross sections computed from KCBK fit compared with the light quark pseudodata data and HERA reduced cross section data~\cite{Aaron:2009aa}. Balitsky + smallest dipole running coupling is used with $\Yobk = \ln 1/0.01$.}
        \label{fig:sigmar-lightq-x1-pd}
\end{figure}

Next we consider fits to our interpolated light quark data set. The fit results are also shown in Tables.~\ref{tab:fits-kcbk}, \ref{tab:fits-resumbk} and~\ref{tab:fits-trbk}. 
Figure~\ref{fig:sigmar-lightq-x1-pd} shows a comparison between the HERA and interpolated light quark data with one of the fits, obtained with the KCBK equation with the Balitsky + smallest dipole running coupling and initial condition parametrized at $\Yobk=\ln 1/0.01$.

The light quark only fits have quite distinct systematics in comparison to the actual HERA data fits. Every single fit setup used needs a substantially larger $C^2$ and to a varying degree larger anomalous dimensions. Lastly, and importantly, light quark fits need larger values of $\sigma_0$ compared to the corresponding total HERA cross section fit.

The slow evolution speed (visible as a large $C^2$ especially when using the parent dipole prescription) and a large $\sigma_0$ in the light quark pseudodata fits can be understood to result from an effective description of non-perturbative effects. We expect that there is a  non-perturbative hadronic contribution in the light quark production cross section which is large (resulting in a large $\sigma_0$) and evolves more slowly as a function of Bjorken-$\xbj$ than  the fully perturbative cross sections, like charm production. In our framework, these non-perturbative effects correspond to large dipoles, with sizes larger than roughly the inverse pion mass. In this case, quark-antiquark dipoles are not the right degrees of freedom, and one should in principle use an another effective desription for the non-perturbative physics, e.g. the vector meson dominance~\cite{Gribov:1968gs,Brodsky:1969iz,Ritson:1970yu,Sakurai:1972wk} model.  

The same non-perturbative effects are there also in the total reduced cross section, and consequently in our fits to full HERA data. However, the full reduced cross section also includes the more reliably perturbative charm production contribution (and a small b quark  one), with a much faster $x$ evolution and a smaller magnitude ($\sigma_0$). Consequently, when performing our (massless) NLO fits to the full HERA data more weight is given to perturbative contributions compared to light quark fits, and there is less need for the fit parameters to adjust to nonperturbative effects with unnatural values.

These observations are compatible with some of the previous analyses. In the study by the AAMQS collaboration~\cite{Albacete:2010sy} it was found that a combined fit to both charm and total reduced cross section requires one to introduce separate fit parameters for the charm quarks, especially the charm quarks require a smaller $\sigma_0$.  A slowly evolving non-perturbative contribution to the light quark production  was also found to be necessary in Refs.~\cite{Berger:2011ew, Mantysaari:2018zdd}.
In the dipole picture applied here, one finds that very large dipoles up to a few femtometers contribute significantly to the light quark structure function~\cite{Mantysaari:2018nng}. In reality, non-perturbative confinement scale effects not included in our perturbative calculation are expected to dominate in these cases as discussed above.

To arrive at one of our central points of this article, we make the observation that even though the HERA DIS data has been described well with leading order dipole picture fits with the BK equation in the past, simultaneous fits to the full data and charm quark data have not been successful with a single BK-evolved amplitude (note however the existence of fits~\cite{Kowalski:2006hc,Rezaeian:2013tka,Soyez:2007kg} using parametrizations that mimic BK evolution). Similar results are found in the recent study with the  target rapidity BK prescription as well \cite{Ducloue:2019jmy}: fits to the full data are excellent but the fit parametrizations do not describe the heavy quark data. Our next-to-leading order analysis, where we separately consider the light quark production only, results in similar  conclusions. This indicates that the description of the light quark contribution has a large theoretical uncertainty as well in any such fit to the full DIS data.

Thus we find that it would be preferable to fit the charm quark structure function $F_{2,c}$ separately (or inclusive $F_L$ data, as the longitudinal photon splits generally to smaller dipoles, resulting in smaller non-perturbative contributions). The $F_L$ measurements from HERA~\cite{Andreev:2013vha} are however not precise enough for our purposes (see the next section). Very precise $F_L$ data (among with inclusive and charm structure functions) can be expected from the future Electron Ion Collider~\cite{Accardi:2012qut,Aschenauer:2017jsk} or from the LHeC~\cite{AbelleiraFernandez:2012cc}.

\subsection{Beyond HERA}
\label{sec:comparebk}
    \begin{figure}%
    % \centering
        \subfloat[Structure function $F_2$]{
            \includegraphics[width=0.48\textwidth]{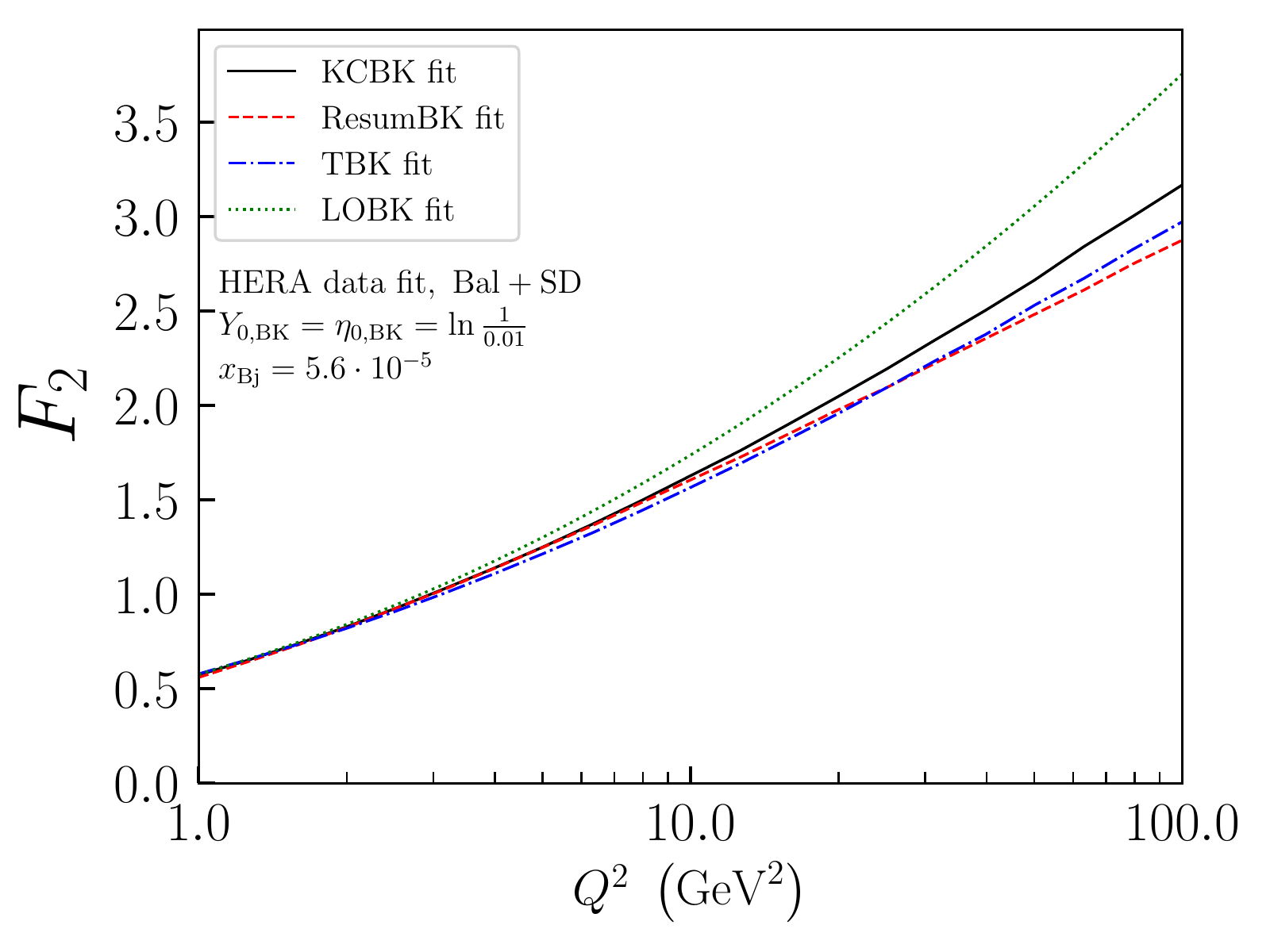}
            \label{fig:lhec-x001-pd-f2}
        }
        \qquad
        \subfloat[Structure function $F_L$]{
            \includegraphics[width=0.48\textwidth]{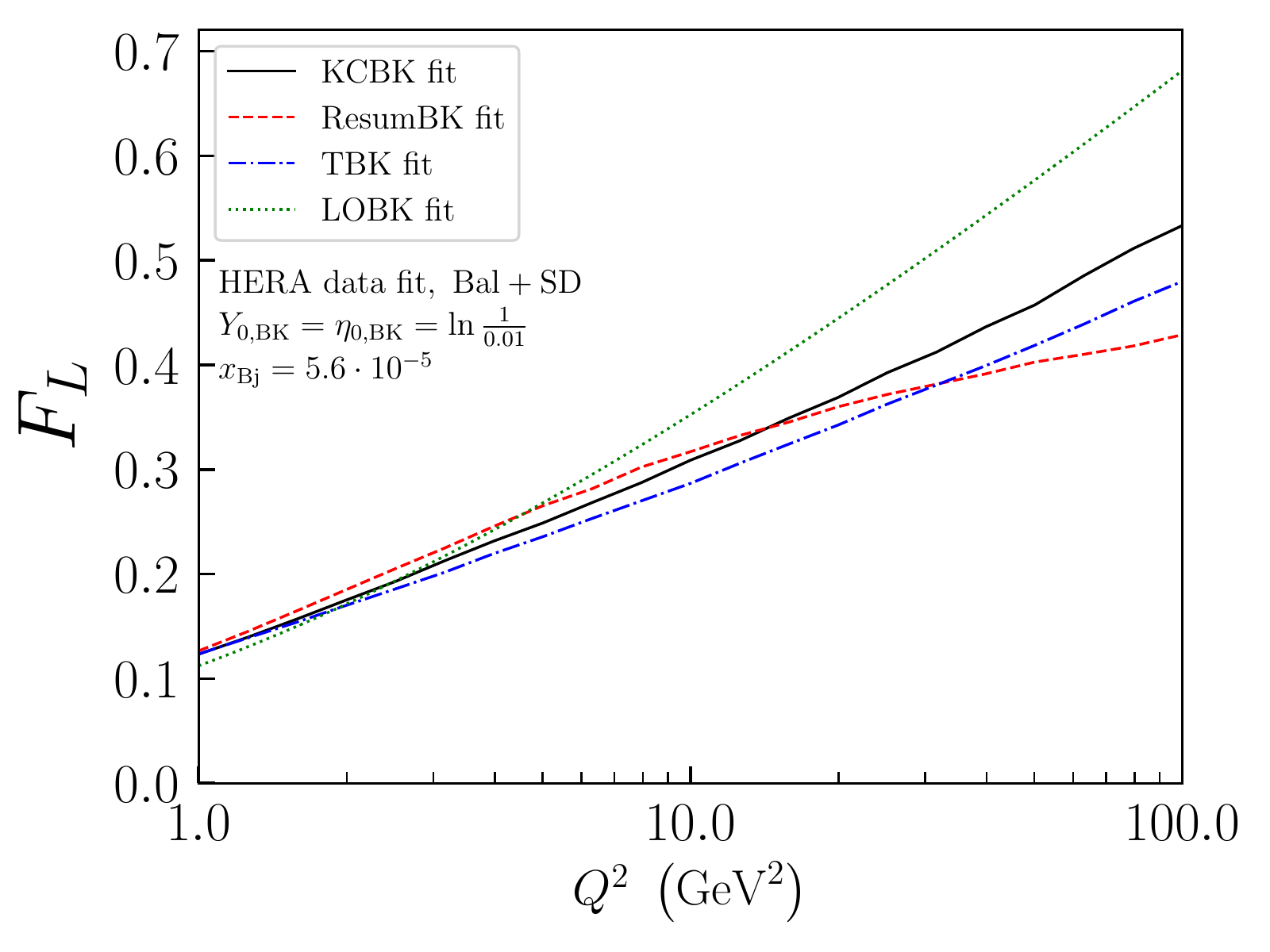}
            \label{fig:lhec-x001-pd-fl}
        }
    \caption{Structure functions $F_2$ and $F_L$ computed from HERA data fit parametrizations extrapolated to the LHeC kinematics. For comparison, the corresponding leading order predictions from Ref.~\cite{Lappi:2013zma} are shown. 
    }
    \label{fig:lhec-f2-fl}
    \end{figure}

Given the equality in the capabilities of the different versions of the BK equation in describing the HERA and light quark data, a question arises if it is possible to distinguish the different fit schemes and find the preferred form of the BK equation. In general, one might expect to see differences in the $Q^2$ dependence of the structure functions at small $x$ (in the HERA kinematics, the fit procedure ensures a compatible evolution). This is because the $Q^2$ dependence is controlled by the anomalous dimension, which behaves differently in ResumBK and KCBK evolutions, when compared to the BK equation formulated in the target momentum fraction as shown in Figs.~\ref{fig:anomdim-kcbk-x001} and \ref{fig:anomdim-trbk-x001}.

At asymptotically small $x$ both approaches can result in the same $Q^2$ dependence of the cross section in spite of the different anomalous dimensions. This can be seen as follows. Let us first consider the BK equation formulated in the target rapidity, and write the dipole amplitude as $N \sim (Q_s^2 r^2)^\gamma.$ The TBK equation results in the saturation scale scaling as $Q_s^2\sim x^{-\lambda}$, as the evolution range is $\ln 1/\xbj$. This gives $N \sim \xbj^{-\lambda \gamma} (Q^2)^{-\gamma}$, and consequently the structure functions behave as
\begin{equation}
    \frac{1}{|\psi^{\gamma^* \to q\bar q}|^2} F_{2,L}(Q^2) \sim (Q^2)^{-\gamma}
\end{equation}
where we have scaled out the $Q^2$ dependence originating from the virtual photon wave function $\psi^{\gamma^* \to q\bar q}$. Substituting an asymptotic anomalous dimension $\gamma\sim 0.7$ we get
\begin{equation}
\label{eq:f2fl_q2_trbk}
    \frac{1}{|\psi^{\gamma^* \to q\bar q}|^2}  F_{2,L}(Q^2) \sim (Q^2)^{-0.7}
\end{equation}

On the other hand, when applying the KCBK or ResumBK equations formulated in terms of the projectile momentum fraction, the evolution range is controlled by $\ln W^2 = \ln (Q^2/\xbj)$. Consequently, we get $Q_s^2 \sim (W^2)^\lambda\sim (Q^2/\xbj)^\lambda $. This gives
\begin{equation}
     \frac{1}{|\psi^{\gamma^* \to q\bar q}|^2} F_{2,L} \sim (Q^2)^{\gamma(\lambda-1)}.
\end{equation}
In general, in the case of ResumBK and KCBK we expect $\gamma \sim 1$ as the evolution does not change the asymptotic anomalous dimension. Using  $\lambda\sim 0.3$ for the generic evolution speed we get
\begin{equation}
    F_{2,L} \sim |\psi^{\gamma^* \to q\bar q}|^2 (Q^2)^{-0.7}
\end{equation}
which is the same $Q^2$ scaling as obtained in case of TBK equation, see Eq.~\eqref{eq:f2fl_q2_trbk}.

In practice, however, in  HERA or even LHeC kinematics the TBK evolution has not reached its asymptotic form, and the anomalous dimension is still close to unity as shown in Fig.~\ref{fig:anomdim-trbk-x001}. Consequently, the $Q^2$-dependence is expected to be  slower in the TBK evolution in realistic kinematics. We note that the structure functions are not actually sensitive to the slope of the dipole at asymptotically small $r$ but in the region $r\sim 1/Q_s$ or $r\sim 1/Q
$, which makes it in practice difficult to compare $Q^2$ dependences analytically. We also note that when computing the structure function at low $\xbj$, also dipole amplitudes at higher $\xbj$ are probed when performing the $z_2$ integral. 

The numerically calculated $Q^2$ dependence of the structure functions $F_2$ and $F_L$ is shown in  Fig.~\ref{fig:lhec-f2-fl}. The results are shown at small $\xbj=5.6 \cdot 10^{-5}$ corresponding to the LHeC kinematics using each of the BK equations, employing the fit to the full HERA data with the Bal+SD running coupling prescription and $\Yobk= \etaobk =\ln \frac{1}{0.01}$.
For comparison, the leading order result based on Ref.~\cite{Lappi:2013zma} is shown. 
Compared to the leading order fit, the $Q^2$ dependence is weaker at next-to-leading order, due to the different asymptotic shape of the dipole amplitude (the leading order BK equation develops a small anomalous dimension $\gamma$ which results in faster $Q^2$ dependence).

The different fit schemes that result in an equally good description of the HERA data start to differ slightly at large $Q^2$ when considering the Bjorken-$\xbj$ region not included in the fits. The longitudinal structure function $F_L$ is more sensitive to small dipole sizes, and as such it can be expected to be more sensitive on the details of the evolution. This is especially visible when the ResumBK evolution is compared to other approaches: the $Q^2$ dependence is much weaker at large $Q^2$. This is due to the resummation of single transverse logarithms not included in other evolution schemes, which has the largest effect at small parent dipole sizes probed at large $Q^2$. 
However, in the realistic kinematical range considered here, the difference between the fits is moderate. This suggests that our next-to-leading order predictions for the structure functions in the future collider experiments are robust.
Future high energy DIS data from e.g. LHeC will be extremely precise, with the expected uncertainties in the structure function measurements being even at the per mill level~\cite{AbelleiraFernandez:2012cc}. As such one could be sensitive to details in NLO BK evolution, even though the effects are not large. Ultimately more differential measurements in addition to the reduced cross section will be needed.

    \begin{figure}[tb]
    % \centering
            \includegraphics[width=0.4726\textwidth]{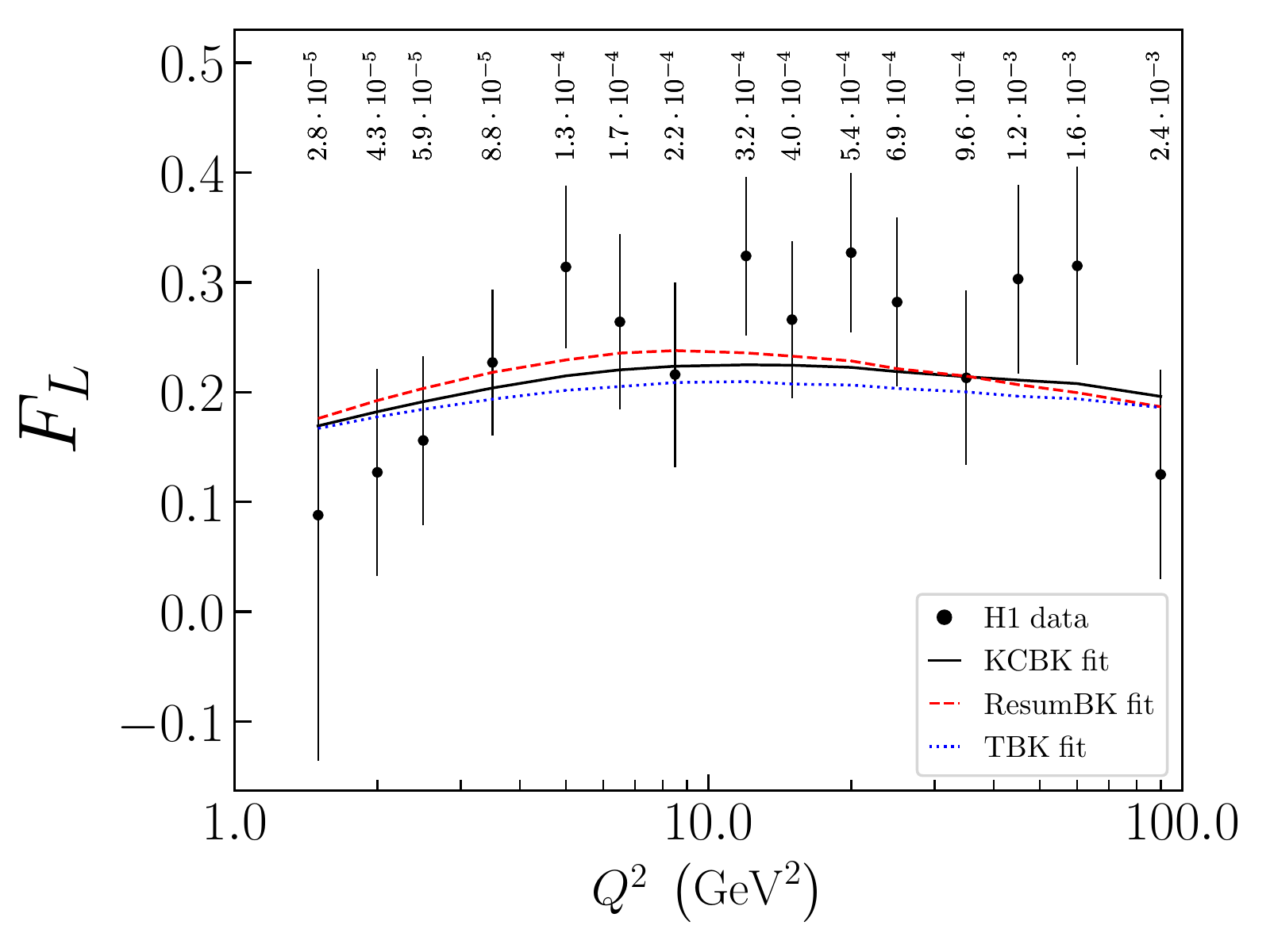}
                    \caption{
                    NLO fit predictions of $F_L$ compared to averaged H1 data \cite{Andreev:2013vha}. Fits are to HERA data, with smallest dipole coupling and $\Yobk = \etaobk= \ln \frac{1}{0.01}$.
                    }
            \label{fig:h1-fl-x001}
    \end{figure}

The most precise measurement of the proton longitudinal structure function $F_L$  up to date is performed by the H1 collaboration at HERA~\cite{Andreev:2013vha} (with compatible results obtained by the ZEUS collaboration~\cite{Abramowicz:2014jak}). 
In Fig.~\ref{fig:h1-fl-x001} we compare the $F_L$ computed from our fits to the H1 data.
Due to the limited statistics, the most precise results are not reported as a function of both $x$ and $Q^2$, but at fixed $x,Q^2$ combinations. Consequently, it is crucial to note that the higher $Q^2$ points are measured at higher $x$. 
All three fit setups result in almost identical $F_L$, as expected as the $F_L$ is measured in the kinematical domain mostly included in our reduced cross section fits. Even though the future Electron Ion Collider~\cite{Accardi:2012qut} will not reach as small Bjorken-$x$ values as the LHeC, the $F_L$ measurements it can perform will be very useful as the HERA measurement has large uncertainties and it only covers a small fraction of the phase space where the details of the evolution can not be accessed.

\section{Conclusions}

We have performed, for the first time, a fit to the HERA structure function data in the Color Glass Condensate framework at next to leading order accuracy in the case of massless quarks. As the full next to leading order BK equation is computationally demanding, we approximate it by employing evolution equations that resum higher order corrections enhanced by large transverse logarithms. As a result of the fits, we obtain the initial condition for the perturbative BK evolution. The resulting dipole-target scattering amplitude can be used in other phenomenological applications, for example when calculating particle production in proton-nucleus collisions at next to leading order in $\as$. 

Similarly as in the leading order fits previously studied in the literature, we find that it is possible to obtain an excellent description of the precise combined HERA structure function data. Equally good fits are obtained when using both the BK equation formulated in terms of the projectile momentum fraction, and the recently proposed BK equation where the evolution rapidity is dictated by the fraction of the target longitudinal momentum.  When extrapolated to LHeC energies, the different BK evolution prescriptions are found to result in moderate differences in the $Q^2$ dependence of the structure functions. This suggests that the NLO calculation presented here is robust, and has a strong predictive power for future DIS measurements in new experimental facilities such as the EIC or LHeC.

As next-to-leading order impact factors for massive quarks are not yet available, it is not possible to compute charm and bottom contribution to the structure functions. To perform consistent fits, we also generated an interpolated light quark data set by subtracting the interpolated charm and bottom contribution from the HERA reduced cross section data. Fits to this light quark data require a much slower Bjorken-$x$ evolution than we naturally get from the perturbative evolution equations applied. Additionally, the apparent proton transverse size obtained is significantly larger than seen when fitting the full HERA data. These features we interpret to result from a non-perturbative hadronic component in the light quark production cross section. This component is large (resulting in a large proton transverse area), and evolves more slowly as a function of Bjorken-$x$, as expected for a hadronic component.

Our results demonstrate the need for massive quark impact factors at next-to-leading order accuracy in the CGC framework, which would allow fits to fully perturbative charm cross section 
separately. Precise measurements of the charm structure function over a wide range of $x$ and $Q^2$, in addition to the longitudinal structure function, from future experiments will also be useful. 
The fits to the generated light quark data  should in principle be considered our principal preferred fits 
as there the agreement between the data and the massless theory should be on the most solid footing. However, if used for QCD phenomenology in other observables where the presumed non-perturbative contribution is smaller, the best one can do is use the full HERA data fits.

In addition to inclusion of the quark masses and the usage of the full NLO BK, the NLO DIS calculation can be improved by relaxing some of the kinematical assumptions. First, in addition to the gluon momentum fraction $z_2$, the quark momentum fraction $z_1$ should not be allowed to get arbitrary close to endpoints $z_1\to 0,1$ in order to avoid production of $q\bar q$ pairs with invariant mass larger than the center-of-mass energy. Additionally, in the virtual correction one should also perform the integral over the gluon momentum fraction and evaluate the dipole operator at the same rapidity as in the real term. This would make it possible to also consistently include a $Q^2$ dependent evolution range in the virtual contribution. Finally, when the Balitsky prescription for the running coupling is used in the BK evolution, there is a mismatch in the running coupling schemes between the impact factor and the evolution equation which could be improved. We plan to address these issues in  future work.

\section*{Acknowledgements}
We thank B. Ducloué for discussions.
This work was supported by the Academy of Finland, projects 314764 (H.M) and  321840 (T.L). G.B, H.H and T.L are supported  under the European Union’s Horizon 2020 research and innovation programme by the European Research Council (ERC, grant agreement No. ERC-2015-CoG-681707) and by the STRONG-2020 project (grant agreement No 824093). The content of this article does not reflect the official opinion of the European Union and responsibility for the information and views expressed therein lies entirely with the authors. Computing resources from CSC – IT Center for Science in Espoo, Finland and from the Finnish Grid and Cloud Infrastructure (persistent identifier \texttt{urn:nbn:fi:research-infras-2016072533}) were used in this work.

\bibliographystyle{JHEP-2modlong.bst}
\bibliography{refs}

\end{document}